\documentclass[12pt]{article}
\usepackage{authblk}
\usepackage[bookmarksnumbered, colorlinks, plainpages]{hyperref}
\usepackage{amsmath, amsthm, amscd, amsfonts, amssymb, graphicx, color, booktabs,tikz}
\usepackage{cite}
\usepackage{fancyhdr}
\usepackage{xspace}
\usepackage{subfigure}
\usetikzlibrary{patterns}
\usetikzlibrary{decorations.pathmorphing}
\usetikzlibrary{decorations.markings}
\usetikzlibrary{arrows.meta}
\tikzset{->-/.style={decoration={
  markings,
  mark=at position #1 with {\arrow{stealth}}},postaction={decorate}}}

\numberwithin{equation}{section}
\definecolor{email}{rgb}{0.00,0.00,0.84}
\begin{document}
\setcounter{page}{1}

%bra ket notation
\newcommand{\bra}[1]{\langle #1 \rvert}
\newcommand{\ket}[1]{\lvert #1 \rangle}

\def\Bbar {\kern 0.18em\overline{\kern -0.18em B}{}\xspace}

\title{\boldmath \large \bf 12th Workshop on the CKM Unitarity Triangle\\ Santiago de Compostela, 18-22 September 2023 \\ \vspace{0.3cm}
\LARGE Summary of Working Group 4: Mixing and
mixing-related $CP$ violation in the B system: 
$\Delta m$, $\Delta \Gamma$, $\phi_s$, $\phi_1/\alpha$, $\phi_2/\beta$, $\phi_3/\gamma$}

\author[1]{Agnieszka Dziurda}
\author[2]{Felix Erben}
\author[3]{Marc Quentin Führing}
\author[4]{Thibaud Humair}
\author[5]{Thomas Latham}
\author[6]{Eleftheria Malami}
\author[7]{Pascal Reeck}
\author[6]{Vladyslav Shtabovenko}
\author[8]{Melissa Cruz Torres}
\author[9]{Yuma Uematsu}
\author[10]{K. Keri Vos}

%PR: I'm not sure if the intent was to sort this alphabetically; in any case I don't mind at all 

\affil[1]{Institute of Nuclear Physics Polish Academy of Science, Poland}
\affil[2]{CERN, Theoretical Physics Department, Geneva, Switzerland}
\affil[3]{TU Dortmund, Germany}
\affil[4]{DESY, Germany}
\affil[5]{University of Warwick, Coventry, United Kingdom}
\affil[6]{University of Siegen, Germany}
\affil[7]{Karlsruhe Institute of Technology}
\affil[8]{National Autonomous University of Honduras, Honduras}
\affil[9]{High Energy Accelerator Research Organization (KEK), Tsukuba, Japan}
\affil[10]{{Gravitational Waves and Fundamental Physics (GWFP), Maastricht University, Duboisdomein 30, NL-6229 GT Maastricht, the Netherlands}}

%\dedicatory{This paper is dedicated to Professor ABCD}
\maketitle
\fancypagestyle{firststyle}
{
\renewcommand{\headrulewidth}{0pt}
\fancyhead{} % clear all header fields
\fancyhead[LE,LO]{P3H-24-017, TTP24-006, CERN-TH-2024-043}
}
\thispagestyle{firststyle}
\begin{abstract}
This summary reviews contributions to the CKM 2023 workshop in Working Group 4: mixing and mixing-related $CP$ violation in B system. The theoretical and experimental progress is discussed.
\end{abstract} \maketitle

\tableofcontents

\section{Introduction}

The study of neutral $B$-meson oscillations provides many insights into quark flavour dynamics. The oscillation frequencies can be calculated precisely within the Standard Model (SM), and any experimentally measured deviation could hint at contributions from New Physics (NP). In addition, the study of mixing-induced $CP$ violation gives 
access to angles of the $B^0_s$ and $B^0_d$ Unitary Triangles (UT) which, in combination with independent measurement, provide a stringent test of the flavour structure of the SM.

These proceedings make a summary of contributions from Working Group 4 at the CKM workshop in 2023. In particular, updates on SU(3)-breaking ratio and bag parameters, improvements in determination of $\Delta \Gamma$, and the Cabbibo-Kobayashi-Maskawa (CKM) angle $\phi_3/\gamma$ as well as measurements of $\phi_s$ and $\sin2\beta$ parameters are discussed.

\section{Standard Model predictions of meson-mixing parameters}

\begin{figure} [hbt!]
\centering
\includegraphics[width=0.45\textwidth]{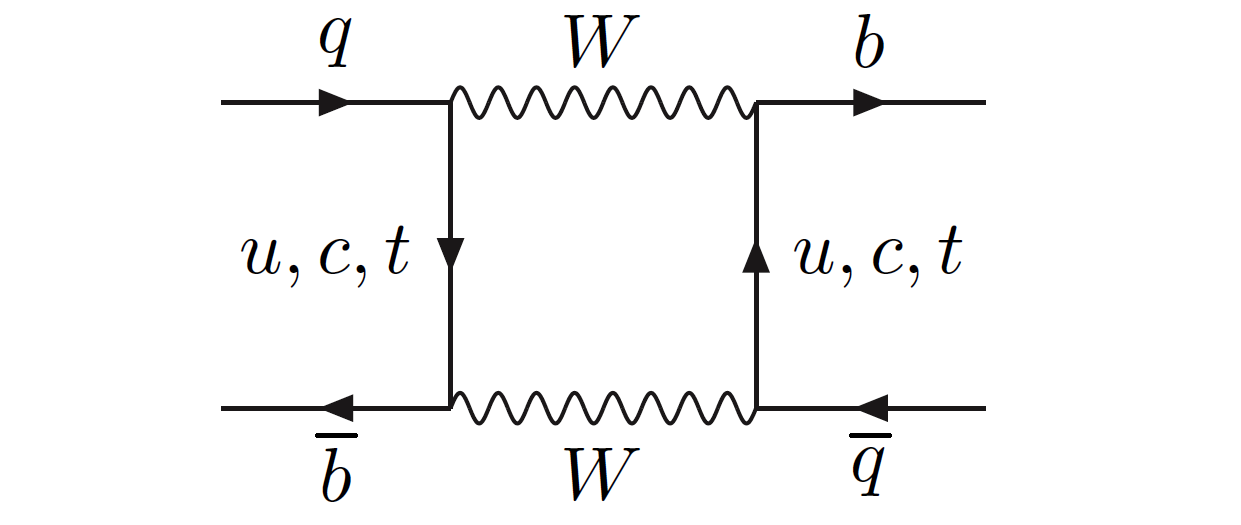}
\includegraphics[width=0.45\textwidth]{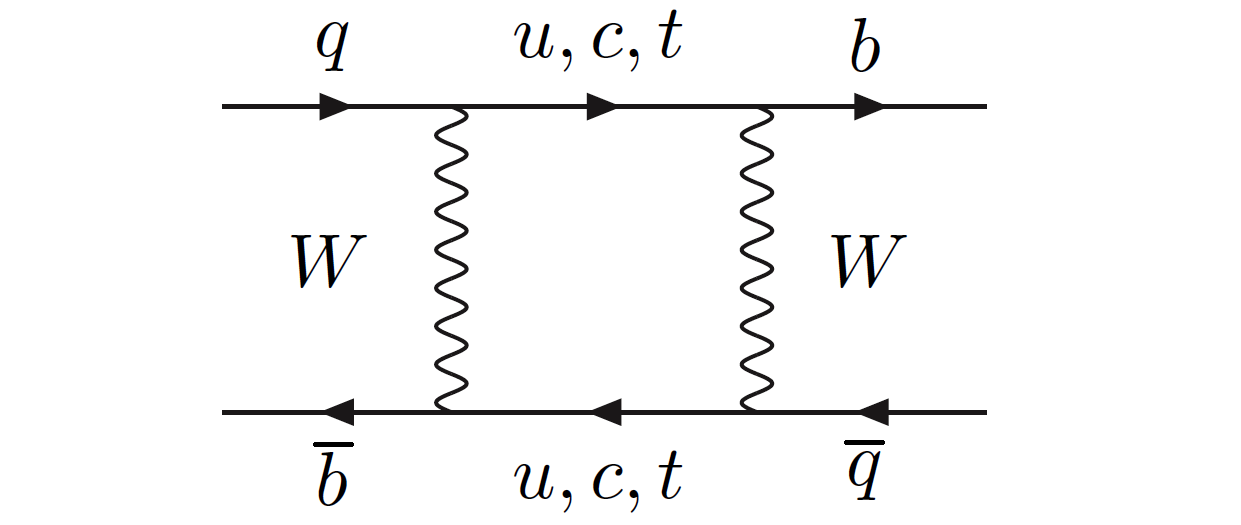}
 \caption{SM diagrams for the $B_q^0-\bar{B}_q^0$ mixing}
\label{fig:Fig1}
\end{figure}

\noindent In the SM, neutral mesons mixing arises from a mismatch between the mass and flavour eigenstates describing the meson-antimeson system.
In the $B^0_{q} - \bar{B}^0_{q}$ system, where $q=d,s$, the mixing process is described by a box-diagram with internal up, charm and top quarks as shown in Fig.~\ref{fig:Fig1}.  
The contribution from internal on-shell and off-shell particles are denoted as $\Gamma^q_{12}$ and $M_{12}^q$ parameters. These parameters are related to the mass ($\Delta M_q$) and width ($\Delta \Gamma_q$) difference as follows: 
\begin{equation*}
\Delta M_q \approx 2 |M_{12}^q|, \quad  \Delta \Gamma_q \approx 2 |\Gamma^q_{12}|\cos{\phi^q_{12}}, \quad a^{q}_{sl} \approx  \Bigg| \frac{\Gamma^q_{12}}{M_{12}^q} \Bigg| \sin{\phi^q_{12}}
\end{equation*}
with $\phi^q_{12} = \arg (-M_{12}^q/\Gamma^q_{12})$ and the parameter $a^{q}_{sl}$ being  the semi-leptonic asymmetry. 
The $B^0_{d,s} - \bar{B}^0_{d,s}$ mixing is hence described by the three physical quantities $\Delta M_q$, $\Delta \Gamma^q_{12}$ $a^{q}_{sl}$, that are equally well accessible to experimental measurements and theoretical calculations. 
%These are the width and mass differences between mass eigenstates, 
%$\Delta \Gamma_{d,s}$ and $\Delta m_{d,s}$ respectively, as well as the CP asymmetry in 
%flavor-specific decays, $a^{d,s}_{\textrm{fs}}$. 
%Both experiment and theory are well aware of the importance of 
These observables are important for obtaining a better handle on understanding of the SM and searching for NP effects.  

The decay-width difference $\Delta \Gamma_{d,s}$ would deviate from its predicted value if New Physics enters via light particles with masses below the electroweak scale. On the other hand, heavy (multi-TeV) degrees of freedom would induce a shift in the measured value of $\Delta m_{d,s}$. The most up-to-date experimental measurements for these two quantities yield~\cite{HFLAV:2022esi}
\begin{alignat}{2} 
    \Delta m^{\rm exp}_d &=  (0.5065 \pm 0.0019)
	\; \text{ps$^{-1}$} , \\
    \Delta\Gamma_d^{\rm exp}/\Gamma_d^{\rm exp} &= 0.001 \pm 0.010 \\
	\Delta m^{\rm exp}_s &=  (17.765 \pm 0.006)
	\; \text{ps$^{-1}$} , \\ 
	\Delta \Gamma^{\rm exp}_s  &=  (0.083 \pm 0.005)\;
	\text{ps}^{-1}. %\qquad \hspace*{1.5em}
 \label{eq:exp}
\end{alignat}

The calculation of neutral meson mixing parameters can be addressed in the framework of an effective theory obtained by integrating out heavy degrees of freedom above the $W$ mass scale. 
The resulting effective  Hamiltonian consists of Wilson coefficients (calculable perturbatively)
multiplying nonperturbative matrix elements that can be obtained from the lattice. In particular, this applies to the dimension-6 matrix elements describing $\Delta B =2$ transitions that enter theory predictions for  $\Delta \Gamma_{d,s}$. The transition to the bag parameters is done by normalising those matrix elements by their vacuum saturation values.

The bag parameter $\hat{B}^{(1)}_{B_q}$ related to the matrix element $O_1 = \bar{b}^i_L \gamma^\mu q^i_L \bar{b}^j_L \gamma^\mu q^j_L$ is also necessary for the theory prediction of $\Delta m_{d,s}$ that can be written as
\begin{equation}
  \Delta m_q = \left|{V_{tb} V_{tq}^*}\right|^2 \, \mathcal{K} \, M_{B_q} \,f^2_{B_q} \hat{B}^{(1)}_{B_q}\,,
\end{equation}
where $f_{B_q}$ is the nonperturbative decay constant, while $\mathcal{K}$
can be computed in perturbation theory.

\section{\boldmath Update on SU(3)-breaking ratios and bag parameters for  \texorpdfstring{$B_{(s)}$}{} mesons}

Lattice QCD calculations of $B$-meson mixing parameters are available from ETMC \cite{ETM:2013jap}, Fermilab/MILC~\cite{FermilabLattice:2016ipl}, RBC/UKQCD~\cite{Boyle:2018knm} and HPQCD~\cite{Dowdall:2019bea}. These computations are summarised in the FLAG 21 review~\cite{FlavourLatticeAveragingGroupFLAG:2021npn} as well as a recent review on $B$-meson physics on the lattice \cite{Tsang:2023nay}. There are some small tensions between lattice results, with HPQCD confirming the experimental average of oscillation frequencies $\Delta m_{d,s}$ and Fermilab/MILC confirming it only at the $2 \sigma$ level. The situation is summarised as part of sum-rules calculations of the same quantities \cite{DiLuzio:2019jyq, King:2019lal}. Generally, all available theory predictions of $\Delta m_{d,s}$ are one or two orders of magnitude less precise than the experimental measurements. A current calculation that can help to shed light into these tensions is a joint effort by the RBC/UKQCD and the JLQCD collaborations \cite{Boyle:2021kqn}, aiming to compute $B_{d,s}$ mixing parameters from chiral domain-wall fermions using lattice QCD on a large set of lattice ensembles.

Using lattice QCD, the dominant short-distance contribution to $B_{d,s}$-mixing can be computed nonperturbatively. This $\Delta B=2$ transition is typically parameterised via an operator-product expansion
\begin{align}
    \langle \bar{B}^0_q | \mathcal{H}^{\Delta B=2}_\mathrm{eff} | B^0_q \rangle = \sum_{i=1}^5 C_i \langle \bar{B}^0_q | \mathcal{O}_i | B^0_q \rangle, 
\end{align}
where $| B^0_q \rangle$ is a state interpolating a $B^0_d$ meson ($q=d$) or a $B^0_s$ meson ($q=s$),   $\mathcal{H}^{\Delta B=2}_\mathrm{eff}$ is the effective Hamiltonian, $\mathcal{O}_i$ are the 5 parity-even, dimension-6 $\Delta B=2$ operators and $C_i$ are the corresponding Wilson coefficients. By computing lattice-QCD two-point functions, masses $M_{B_q}$ and decay constants $f_{B_q}$ can be extracted via
\begin{align}
    \langle B_q(t) B_q^\dagger(0) \rangle_{L,a,m_l,m_h} \Rightarrow M_{B_q}(L,a,m_l,m_h), f_{B_q}(L,a,m_l,m_h) \, ,
\end{align}
all at a given volume $L$, lattice spacing $a$, simulated light-quark mass $m_l$ and heavy-quark mass $m_h$. From three-point functions
\begin{align}
    \langle B_q(\Delta T) \mathcal{O}_i(t) B_q^\dagger(0) \rangle_{L,a,m_l,m_h} \Rightarrow \langle \bar{B}^0_q | \mathcal{O}_i | B^0_q \rangle (L,a,m_l,m_h) \, ,
\end{align}
one can extract the desired matrix elements, or alternatively bag parameters
\begin{align}
    \mathcal{B}_{B_q}^{[i]} = \frac{\langle \bar{B}^0_q | \mathcal{O}_i | B^0_q \rangle }{\langle \bar{B}^0_q | \mathcal{O}_i | B^0_q \rangle_\mathrm{VSA}}
    \, ,
    \label{eq:lat-bag}
\end{align}
which are the ratios of the matrix elements by their vacuum saturation approximation. These extracted quantities on the lattice can then be combined and by taking the relevant limits, e.g.
\begin{align}
    \Delta m_q = | V_{td} V_{tq}^*|^2 \kappa \lim_{a \to 0} \lim_{L \to \infty} \lim_{m_l \to m_l^P} \lim_{m_h \to m_h^P} (M_{B_q}f_{B_q}^2\mathcal{B}^{[1]}_{B_q})(L,a,m_l,m_h)
    \, ,
\end{align}
where $\kappa$ is some known perturbative factor and the CKM matrix elements $V_{tq}$ can be taken from experiment to predict $\Delta m_q$ or the other way round. 

It is currently unfeasible to simulate both at the physical light-quark mass $m_l^P$ and physical heavy-quark mass $m_h^P=m_b$. In the project currenlty pursued by the RBC/UKQCD and the JLQCD collaborations \cite{Boyle:2021kqn}, 15 different ensembles at 6 lattice spacings between $0.044\mathrm{fm}$ and $0.11\mathrm{fm}$ are employed, two of which are at the physical pion mass. The heavy-quarks are simulated fully relativistically, using the domain-wall fermion action also employed for the light quarks. Thanks to the ensembles with the finest lattice spacings, heavy-qaurk masses almost reaching $m_b$ can be simulated. 

\begin{figure} [hbt!]
\centering
\includegraphics[width=0.65\textwidth]{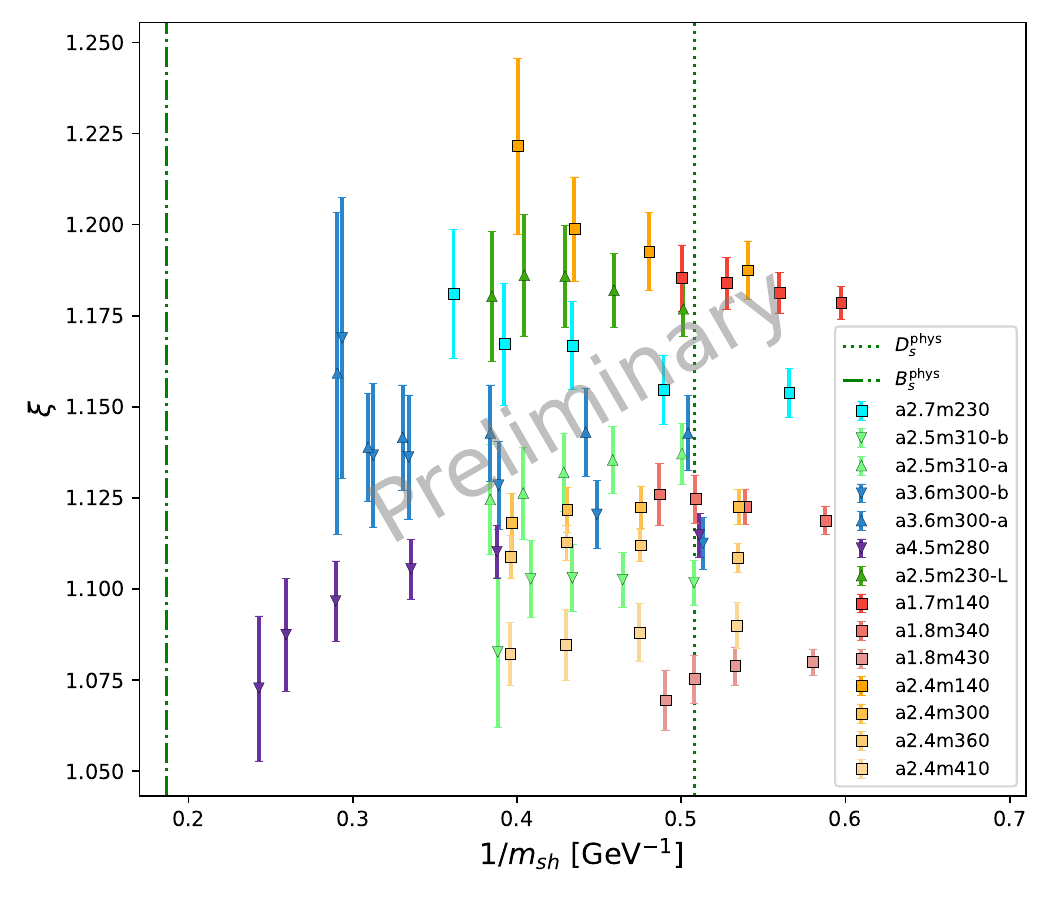} 
\includegraphics[width=0.65\textwidth]{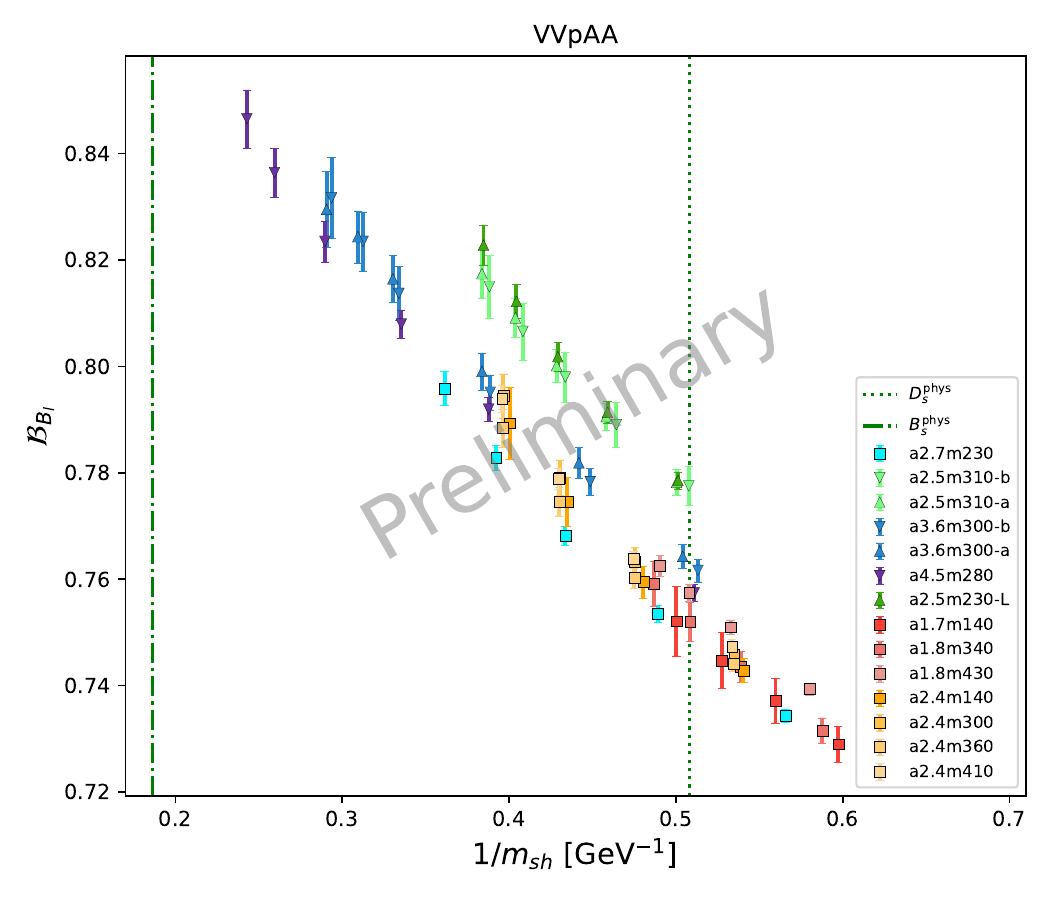}
 \caption{\textbf{Top panel:} The $SU(3)$-breaking ratio $\xi$. \textbf{Bottom panel:} Renormalised bag parameters $\mathcal{B}^{[1]}_{B_d}$ in the RI-SMOM scheme, not yet matched to a continuum scheme. \textbf{In both panels},  multiple heavy-quark masses are shown per ensemble plotted via the inverse heavy-strange meson mass $1/m_{sh}$. Ensemble names indicate the lattice spacing in GeV and pion mass in MeV - e.g. ensemble ``a1.7m14'' has an inverse lattice spacing of $a^{-1}\sim 1.7$GeV and a pion mass of $m_\pi \sim 140$MeV. The vertical green lines indicate the inverse physical $B_s$ mass (left) and $D_s$ mass (right). The RBC/UKQCD ensembles already used in \cite{Boyle:2018knm} have square symbols and the set of JLQCD ensembles has triangle symbols.}
\label{fig:FigLat1}
\end{figure}

The bare bag parameters from Eq.~\eqref{eq:lat-bag} or the bare matrix elements $\langle \bar{B}^0_q | \mathcal{O}_i | B^0_q \rangle$ obtained from lattice QCD need to be renormalised and matched to a continuum scheme to be compared with experiment. This project uses a nonperturbative renomalisation in the so-called RI-SMOM scheme \cite{Boyle:2017skn}. These renormalisation constants cancel in certain ratios such as
\begin{align}
    \xi = \frac{f_{B_s}\sqrt{\mathcal{B}^{[1]}_{B_s}}}{f_{B_d}\sqrt{\mathcal{B}^{[1]}_{B_d}}} 
    \, .
\end{align}
 Figure~\ref{fig:FigLat1} shows preliminary results of the RBC/UKQCD lattice QCD project both on $\xi$ as well as renormalised bag parameters $\mathcal{B}^{[1]}_{B_d}$ in the RI-SMOM scheme for all ensembles and heavy-quark masses employed in this work. In particular, the very fine JLQCD ensembles lead to an increased reach in the simulated heavy-quark masses towards the physical $b$-quark mass, the extrapolation of which was one of the dominating sources of uncertainty in the chiral-continuum fit of the RBC/UKQCD ensembles only~\cite{Boyle:2018knm}. 

\newpage

\section{Mixing angles}

Neutral $B$-meson decays with flavour-changing neutral current
are excellent to probe New Physics contributions, potentially occurring via loop
(penguin) processes. %The comparison of experimental results with theoretical prediction is one of the most powerful SM test. 

\begin{figure} [hbt!]
\centering
\begin{tikzpicture}
		\draw[thick,-{Latex[length=3mm]}] (-0.2,0) -- (6,0) node[anchor=north west] {$\bar{\rho}$};
		\draw[thick,-{Latex[length=3mm]}] (0,-0.2) -- (0,3) node[anchor=north east] {$\bar{\eta}$};
		\draw[blue,thick,-{Latex[length=3mm]}] (0,0) node[anchor=south west,xshift=0.7cm,black] {$\gamma = \varphi_3$}  -- node[above,anchor=south east] {$R_u$}  (3,2) ;
		\draw[black!50!green,thick,-{Latex[length=3mm]}] (5,0) node[anchor=south east,xshift=-0.4cm,black] {$\beta = \varphi_1$} --  (0,0);
		\draw[red,thick,-{Latex[length=3mm]}] (3,2) node[anchor=north,yshift=-0.5cm,black] {$\alpha = \varphi_2$} -- node[above,anchor=south west] {$R_t$}  (5,0);
\end{tikzpicture}
 \caption{CKM unitarity triangle in the $\bar{\rho} - \bar{\eta}$ plane.}
\label{fig:rho-eta-plane}
\end{figure}
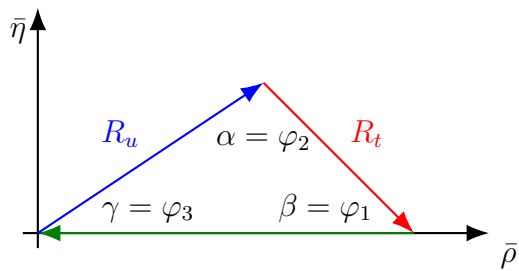

The unitarity triangle with the largest area is defined in the complex plane as
\begin{equation}
V^\ast_{ud} V_{ub} + V^\ast_{cd} V_{cb} + V^\ast_{td} V_{tb} = 0.
\end{equation}
Division of this equation by $V_{cb}^\ast V_{cd}$ fixes the base
of the triangle to have unit length, while its side lengths are given by
\begin{equation}
R_u = \left |\frac{V_{ub}^\ast V_{ud}}{V_{cb}^\ast V_{cd}} \right |, \quad
R_t = \left |\frac{V_{tb}^\ast V_{td}}{V_{cb}^\ast V_{cd}} \right |,
\end{equation}
as shown in Fig.~\ref{fig:rho-eta-plane}. 

Accordingly, the angles of the triangle are defined as: 
\begin{align}
\beta & =  \varphi_1 = \arg \left (- \frac{{ V^\ast_{cb} V_{cd}}}{{ V^\ast_{tb} V_{td}}} \right), \\
\alpha &= \varphi_2 = \arg \left (- \frac{{ V^\ast_{tb} V_{td}}}{{ V^\ast_{ub} V_{ud}}} \right),  \\
\gamma & = \varphi_3 = \arg \left (- \frac{{ V^\ast_{ub} V_{ud}}}{{V^\ast_{cb} V_{cd}}} \right). \label{eq:gamma}
\end{align}
The $\alpha$, $\beta$, $\gamma$ naming convention is used hereafter.
In the context of $B^0_s - \bar{B}^0_s$ mixing it is also useful to consider the rather flat triangle
\begin{equation}
V^\ast_{us} V_{ub} + V^\ast_{cs} V_{cb} + V^\ast_{ts} V_{tb} = 0,
\end{equation}
with
\begin{equation}
\beta_s  = \arg \left (- \frac{V_{ts} V_{tb}^\ast}{ V_{cs} V_{cb}^\ast} \right).
\label{eq:beta_s}
\end{equation}

Experimentally, the angles $\beta$ and $\beta_s$ are determined from the measurements of time-dependent \textit{CP} asymmetries in the $B^0_{q}$ decays to a $CP$ eigenstate $f$. The asymmetry $\mathcal{A}^{C\!P}$ between the number of initially-produced $B^0_{q}$ and of initially-produced  $\bar B^0_{q}$ mesons is function of the decay-time $t$ and reads\footnote{All equations related to the determination of the mixing phases here neglect \textit{CP} violation in the mixing.}
\begin{equation}
\label{eq:asymmetry}
\mathcal{A}^{C\!P}_f(t) = \frac{S_f\sin{(\Delta m_{q} t)}-C_f\cos(\Delta m_{q} t)}{\cosh(\frac{\Delta \Gamma_{q}}{2} t )+\mathcal{A}^{\Delta\Gamma}\sinh(\frac{\Delta \Gamma_{q}}{2} t )}.
\end{equation}
The parameter $S_f$ is related to the mixing phase as $\phi^f_{q}$ $S=sin(2\phi^f_{q})$. Within the SM, and neglecting penguin pollution effects, $\phi_s^f = - 2 \beta_s$ and $\phi^f=2\beta$. The parameter $C_f$ is $C_f=\frac{1-|\lambda^f|^2}{1+|\lambda^f|^2}$, where $|\lambda^f|$ is the ratio between the $\bar B^0_{q}\to f$ and $B^0_{q} \to f$ decay amplitudes:   
$|\lambda^f|=\left|\frac{\bar A_f}{A_f}\right|$.

%Within the SM, the mixing phase $\phi_s^{c \bar{s} s}$ is then $\phi_s^{c \bar{s} s} = - 2 \beta_s$.

\subsection{Determinations of the  \texorpdfstring{$\phi_s$}{} angle}
%\subsection{Determinations of $\phi^{c\bar{c}s}_s$ from $B_s^0 \to J/\psi K^-K^+$ decays} 

\begin{figure} [hbt!]
\centering
\includegraphics[width=0.75\textwidth]{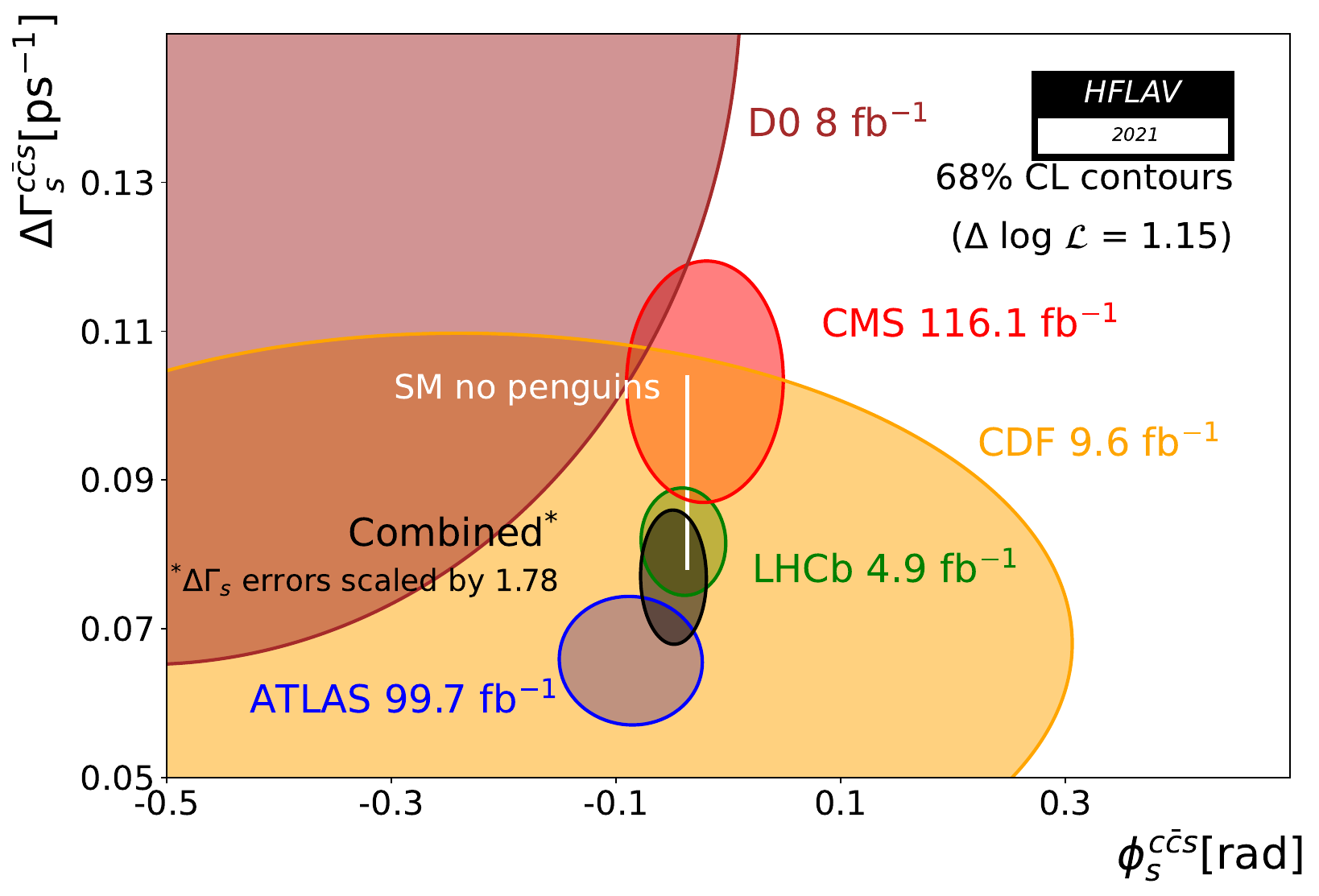}
 \caption{Average of experimental results for $\phi^{c\bar{c}s}_s$ vs $\Delta \Gamma_s$ parameters~\cite{HFLAV:2022esi}.}
\label{fig:phisvsDGs}
\end{figure}

%Within the SM, the phase associated with $B^0_s$-$\bar B^0_s$ mixing is defined in Eq.~\eqref{eq:beta_s}.
 %Any deviations from SM predictions would indicate processes. 
 The ($\phi^{c\bar{c}s}_s$,$\Delta \Gamma_s$) plane with the individual 68\% confidence-level contours of ATLAS, CMS, CDF, D0 and LHCb results for 2021 average is shown in Fig.~\ref{fig:phisvsDGs}.

The golden channel, $B_s^0 \to J/\psi K^-K^+$, has been used in the improved determination of the $\phi^{c\bar{c}s}_s$ parameter based on the 6 fb$^{-1}$ of data collected by the LHCb detector between 2015 and 2018~\cite{LHCb-phis}. This measurement supersedes the previous LHCb analysis~\cite{LHCb-phis-old}. The results are 
\begin{equation*}
\phi^{c\bar{c}s}_s = -0.039 \pm 0.022 \pm 0.006 \quad \rm{rad}, \quad 
|\lambda^{c\bar{c}s}|= 1.001 \pm 0.011 \pm 0.005,
\end{equation*}
finding a good agreement with the SM predictions. No evidence for $CP$ violation is found. Further combination with all LHCb $\phi^{c\bar{c}s}_s$ measurements~\cite{LHCb-phis-run1,LHCb-phis-ee,LHCb-phis-above,LHCb-phis-pipi,LHCb-phis-2S,LHCb-phis-DsDs} 
(from $B_s^0 \to J/\psi K^-K^+$ above $\phi(1020)$ resonance, $B_s^0 \to D_s^-D_s^+$, $B_s^0 \to J/\psi \pi^-\pi^+$, $B_s^0 \to \psi(2S) K^-K^+$ decays), gives 
$\phi^{c\bar{c}s}_s = -0.031 \pm 0.018$. In addition, the analysis provides determination of $\Delta \Gamma_s$ and $\Gamma_s - \Gamma_d$ finding: 
\begin{align*}
\Delta \Gamma_s &=  0.0845 \pm 0.0044 \pm 0.0024\quad\rm{ps}^{-1}, \\
\Gamma_s - \Gamma_d &= -0.0056^{+0.0013}_{-0.0015} \pm 0.0014\quad\rm{ps}^{-1},
\end{align*}
where the first uncertainty is statistical and second is systematic. The obtained values are the most precise up to date and agrees well with SM predictions. 

The CMS experiment reported the measured values of $\phi_s$ and $\Delta \Gamma_s$ to be~\cite{CMS-phis}
\begin{equation*}
\phi^{c\bar{c}s}_s = -11 \pm 50 \pm 0.10 \quad \rm{mrad}, \quad 
\Delta \Gamma_s = 0.114 \pm 0.014 \pm 0.007 \quad \rm{ps}^{-1},
\end{equation*}
using a data sample corresponding to an integrated luminosity of 96.4 fb$^{-1}$ collected in proton-proton collisions at $\sqrt{s}$ = 13 \rm{TeV} in 2017-2018. 

On the other hand, the latest results reported by the ATLAS collaboration~\cite{ATLASDGs} are the following
\begin{align*}
\phi^{c\bar{c}s}_s = -0.087 \pm 0..036 \pm 0.021 \quad \rm{rad},\\
\Delta \Gamma_s = 0.065 \pm 0.0043 \pm 0.0037 \quad \rm{ps}^{-1}, \\
\Gamma_s = 0.6703 \pm 0.0014 \pm 0.0018 \quad \rm{ps}^{-1},
\end{align*}
Obtained by combining the results of 13 \rm{TeV} and 7 and 8 \rm{TeV} of data taking.

%\color{red} Here add CMS \color{black}

%\subsection{Measurement of $\phi^{s\bar{s}s}_s$ from $B_s^0 \to \phi\phi$ decays} 

\begin{figure} [hbt!]
\centering
\includegraphics[width=0.8\textwidth]{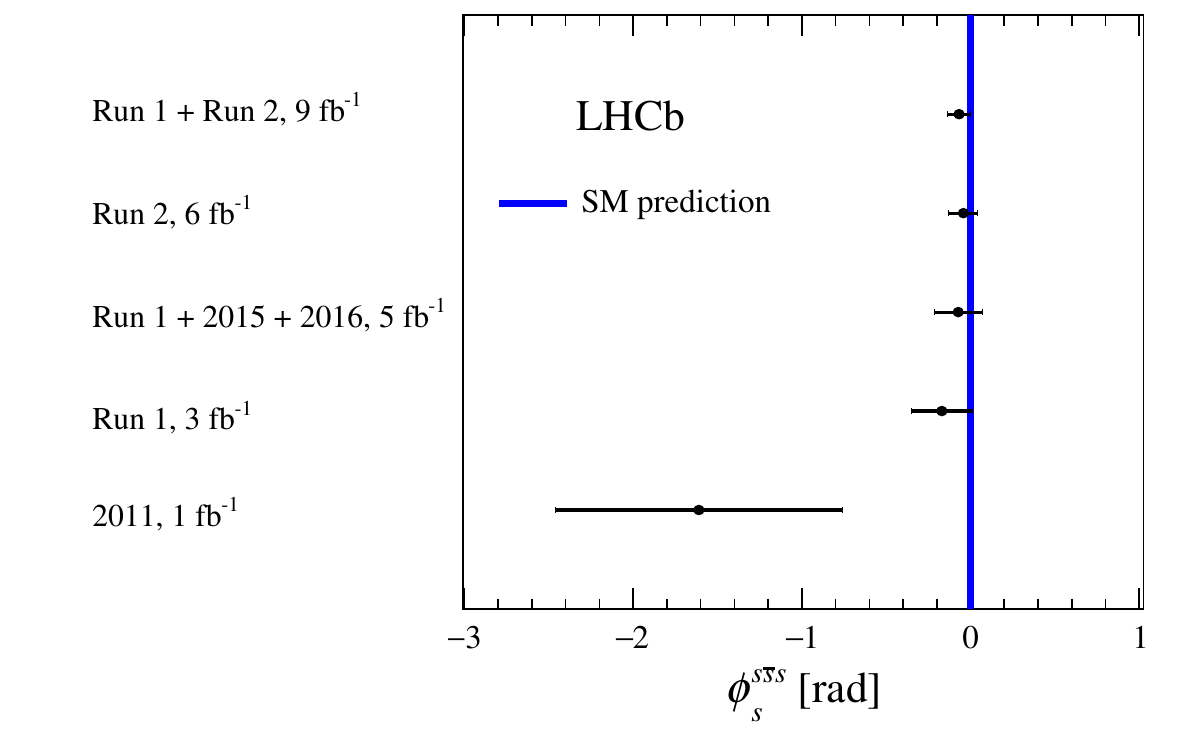}
 \caption{Comparison of $\phi^{s\bar{s}s}_s$ measurements from this and previous analyses by the LHCb collaboration. The vertical band indicates the SM prediction. Run1 stands for 2011-2012 period, while Run2 denotes 2015-2018 years~\cite{LHCb-phiphi}.}
\label{fig:lhcb-phiphi}
\end{figure}

While $B_s^0 \to J/\psi K^-K^+$ decays are dominated by the tree-level transition, $B_s^0 \to \phi\phi$ modes allow to measure mixing angle via $b \to s\bar{s}s$ processes. Within LHCb experiment this channel is a benchmark mode to study $CP$ violation in flavour-changing neutral currents. 
%In this decay the $CP$ violation arise from interference between direct decay and the decay after $b$-hadron mixing.
In addition to the $\phi^{s\bar{s}s}_s$ mixing angle, the parameter $|\lambda|$ is measured. Due to a cancellation of the mixing and decay
weak phase the former is expected to be zero, while for the latter absence of direct $CP$ asymmetry leads to expected value of unity. Any deviations from these values would be signs of New Physics in these decays. In addition, $B_s^0 \to \phi\phi$ decays probe possible polarisation dependent effects in NP thanks to three linear polarisation states in the $\phi\phi$ system.  
The updated measurement~\cite{LHCb-phiphi}  uses the data collected in 2015-2018 which corresponds to 6 fb$^{-1}$ collected by the LHCb detector at the centre-of-mass energy 13 TeV and supersedes previous measurement presented in~\cite{LHCb-phiphi-old} The results are: 
\begin{equation*}
\phi^{s\bar{s}s}_s = -0.042 \pm 0.075 \pm 0.009 \quad \rm{rad}, \quad 
|\lambda|= 1.004 \pm 0.030 \pm 0.009,
\end{equation*}
where the first uncertainty is statistical and second systematic. 
Further combination with the LHCb result from 2011-2012 data~\cite{LHCb-phiphi-run1} yields to $\phi^{s\bar{s}s}_s = -0.074 \pm 0.069$ rad, 
$|\lambda| = 1.009 \pm 0.030$. The comparison with other LHCb analyses is shown in Fig.~\ref{fig:lhcb-phiphi}. The measurement agrees with SM predictions.
In addition, for the first time three polarisations are measured finding a good agreement among them. It is the most precise determination of time dependent $CP$ asymmetry in $B_s^0 \to \phi\phi$  modes and in any penguin-dominated beauty-meson decay.

\boldmath\subsection{\texorpdfstring{$\sin 2\beta$}{} determination at LHCb}\unboldmath 

The CKM angle $\beta$ can be determined from the measurements of time-dependent $C\!P$ asymmetries in the decays of neutral $B$ mesons to final states containing charmonium mesons, namely $B^0 \to \psi K^0_S$ with $\psi = [J/\psi, \psi(2S)]$, which are dominated by $b\to c\bar{c}s$ tree-level transitions.
As the decay-width difference is small in the $B^0$ system, the denominator of Eq.~\eqref{eq:asymmetry} is equal to unity, leaving $C_{\psi K_S}$ and 
$S_{\psi K_S}$ as experimentally measurable observables.
In the SM, $C_{\psi K_S}=0$ and $S_{\psi K_S}=\sin2\beta$, with 
corrections due to penguin pollution expected to be smaller than the 
currently achievable experimental precision.
% $\beta$ as $S = \sin(2\beta + \phi^P + \phi^{N\!P}$), where $\phi^P$ is the possible penguin contribution (suppressed in these decays) and $\phi^{N\!P}$ stands for possible NP contribution.
%Therefore within the SM it is expected that $S = \sin (2\beta)$ and $C = 0$ to within the currently achievable precision. 
Historically, these measurements have been performed by the BaBar, Belle and LHCb collaborations, resulting in a world-average of $\sin (2\beta) = 0.699 \pm 0.017$\cite{HFLAV:2022esi}.

\begin{figure} [hbt!]
\centering
\includegraphics[width=0.8\textwidth]{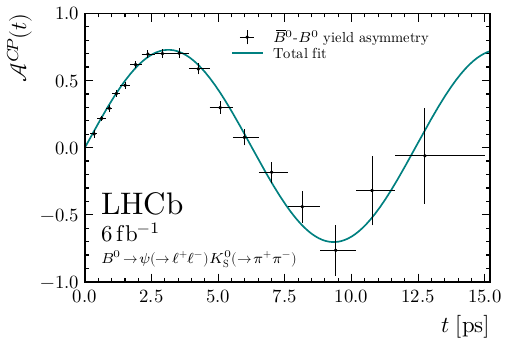}
 \caption{Time-dependent $C\!P$ asymmetry of $B^0 \to J/\psi K^0_S$ decays~\cite{LHCb-2beta}.}
\label{fig:lhcb-sin2b}
\end{figure}

The LHCb collaboration has performed new measurements of $C\!P$-violating parameters from $B^0 \to \psi K^0_S$ decays~\cite{LHCb-2beta} using the full Run2 (2015-2018) data set, corresponding to the integrated luminosity of 6 fb$^{-1}$ collected by the LHCb detector at the centre-of-mass energy $\sqrt{s}=13$ TeV.
Three decay modes are used: $B^0 \to J/\psi (\to \mu^+\mu^-) K^0_S$, $B^0 \to \psi(2S) (\to \mu^+\mu^-) K^0_S$ and $B^0 \to J/\psi (\to e^+e^-) K_S^0$.
In all cases, the $K^0_S$ candidates are reconstructed in the decay $K^0_S \to \pi^+\pi^-$.
Due to the long lifetime of the $K^0_S$, it is necessary to consider a number of different reconstruction categories, depending on whether the pions left hits in the various parts of the LHCb tracking system.
Two such categories are included for the first time in a $C\!P$-violation measurement: that where only one of the pions leaves hits in the vertex locator, and that where one of the pions leaves hits only upstream of the magnet.
Together, these boost the signal yields by approximately 13\%.
The time-dependent asymmetries are shown in Fig.~\ref{fig:lhcb-sin2b}, while the $C\!P$-violating observables are measured to be: 
\begin{align*}
S_{\psi K^0_S}^{\text{Run2}} &= 0.717 \pm 0.013(\text{stat}) \pm 0.008(\text{syst}), \\
C_{\psi K^0_S}^{\text{Run2}} &= 0.008 \pm 0.012(\text{stat}) \pm 0.003(\text{syst}),
\end{align*}
where the first uncertainty is statistical and second is systematic, and a correlation coefficient of 0.441 is found between the two observables.
These are the most precise measurement of these quantities and are more precise than the current world average values~\cite{HFLAV:2022esi}.
The combination with the previous measurement using Run~1 data is performed, resulting in: 
\begin{align*}
S_{\psi K^0_S}^{\text{Run1+2}} &= 0.724 \pm 0.014,\\
C_{\psi K^0_S}^{\text{Run1+2}} &= 0.010 \pm 0.012,
\end{align*}
where statistical and systematic uncertainties are combined and the correlation between observables is 0.40.  
The results agree with the world average~\cite{HFLAV:2022esi} and global fits by CKMFitter~\cite{CKMFitter} and UTFit~\cite{UTFit} collaborations. 

\boldmath\subsection{ \texorpdfstring{$\beta$}{} and related measurements at Belle~II}\unboldmath

The Belle~II collaboration has recently performed several measurements of the mixing-induced $CP$-violating parameters in $B^0\to\eta' K_S^0$, $B^0\to \phi K^0_S$, $B^0\to K^0_S\pi^0$ and $S^{\rm Belle}_{K^0_SK^0_SK^0_S}$ decays. 
These decays are mediated by a $b\to s$ loop-suppressed transition in the 
SM, making them sensitive to NP. These transitions carry approximately the same weak phase as the tree-level $b\to\bar c\bar c s$ transition.
Hence $S \approx \sin2\beta$ in these decays, up to small corrections related to
sub-leading amplitudes. 

\begin{figure} [hbt!]
\centering
\subfigure[$B^0\to K^0_S\pi^0$ decay]{
\includegraphics[height=0.4\textwidth]{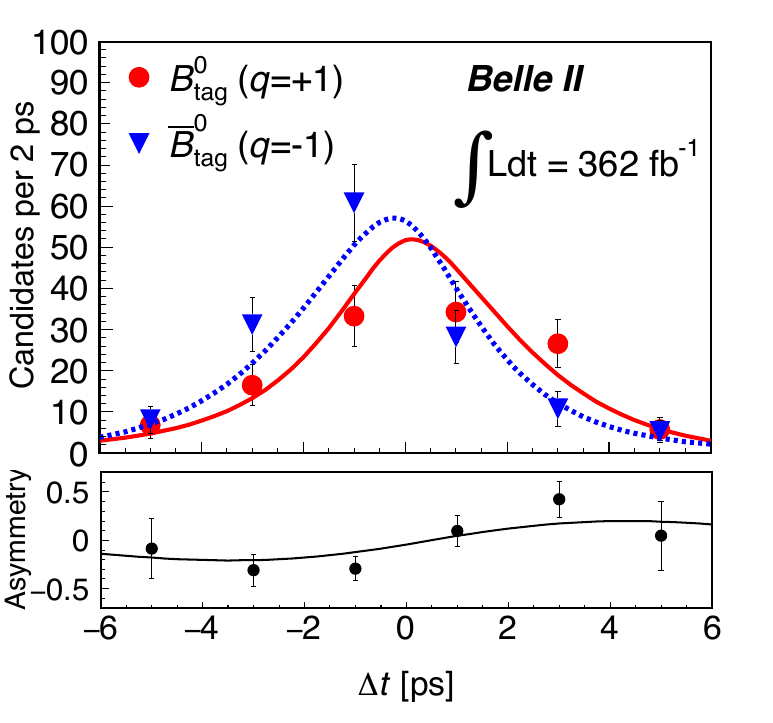}
}
\subfigure[$B^0\to K^0_S\pi^0\gamma$ decay]{
\includegraphics[height=0.4\textwidth]{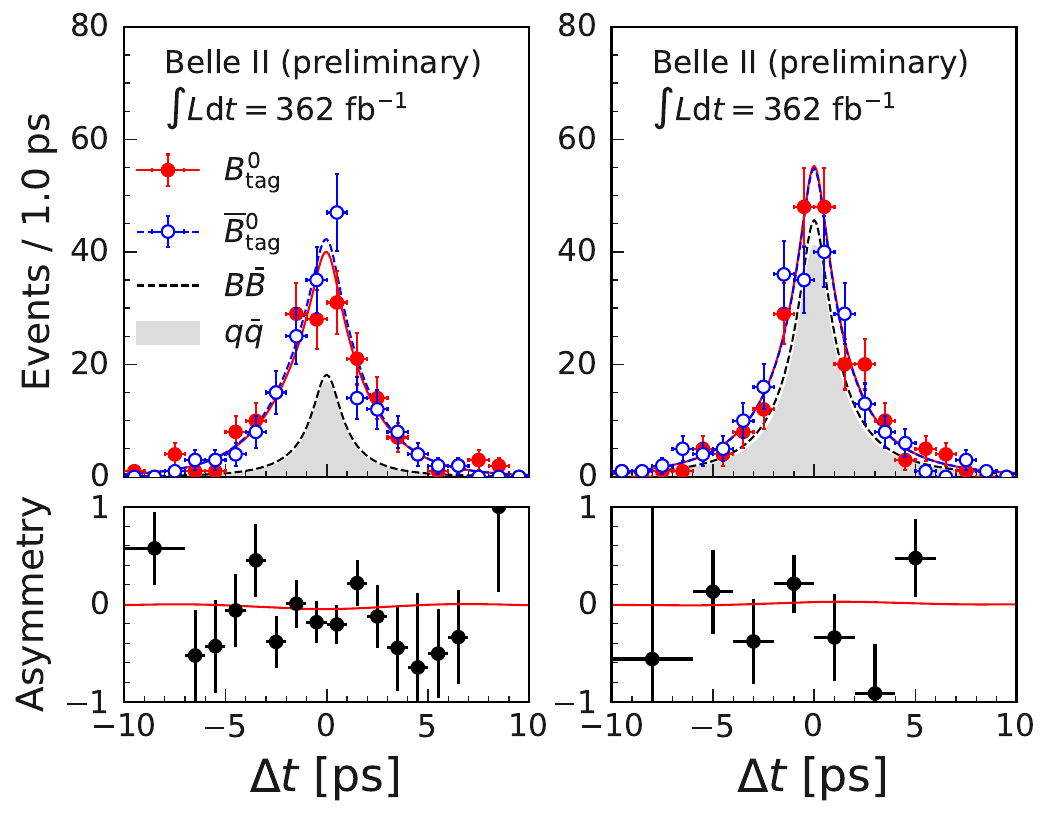}
}
\caption{
Decay-time difference $\Delta t$ distribution in $B^0\to K^0_S\pi^0$ decay~\cite{Belle-II:2023grc} and $B^0\to K^0_S\pi^0\gamma$ decay.
The latter decay is separately analyzed depending on $K^0_S\pi^0$ invariant mass: $m_{K^0_S\pi^0}\in (0.8, 1.0)\,{\rm GeV}\!/c^2$, corresponding to $K^{*0}\to K^0_S\pi^0$ decays; and $m_{K^0_S\pi^0}\in (0.6, 0.8)\cup(1.0, 1.8)\,{\rm GeV}\!/c^2$, corresponding to the non-resonant decays.
$\Delta t$ distribution is shown separately for a flavor of the accompanying $B$ mesons from $\Upsilon(4S)$ decays, $B_{\rm tag}$.
The signal-subtracted distribution is shown for $B^0\to K^0_S\pi^0$ decay.
The asymmetry $[N_{\rm sig}(B^0_{\rm tag})-N_{\rm sig}(\Bbar^0_{\rm tag})]/[N_{\rm sig}(B^0_{\rm tag})+N_{\rm sig}(\Bbar^0_{\rm tag})]$ is shown in the bottom panels.
}
\label{fig:BelleIIdt}
\end{figure}

%The Belle~II collaboration has started to reveal its potential to measure $C\!P$-violating parameters.
%The modern analysis techniques enable measurements competitive with its predecessor, the Belle collaboration, even using a smaller data sample of $(388\pm6)\times10^6\ B\Bbar$ events.
%Using the 
%The latest Belle measurements of $S$ for various decay channels sensitive to NP contributions are
%\begin{align*}
%    S^{\rm Belle}_{\eta'K^0_S}      &= \phantom{-} 0.68 \pm 0.07 \pm 0.03       ~\text{\cite{Belle:2014atq}}, \\
%    S^{\rm Belle}_{K^0_S\pi^0}      &= \phantom{-} 0.67 \pm 0.31 \pm 0.08       ~\text{\cite{Belle:2008kbm}}, \\
%    S^{\rm Belle}_{\phi K^0_S}      &= \phantom{-} 0.50 \pm 0.21 \pm 0.06       ~\text{\cite{Belle:2006dlp}}, \\
%    S^{\rm Belle}_{K^0_SK^0_SK^0_S} &=          -  0.71 \pm 0.23 \pm 0.05       ~\text{\cite{Belle:2020cio}}, \\
%    S^{\rm Belle}_{K^{*0}\gamma}    &=          -  0.32^{+0.36}_{-0.33}\pm 0.05 ~\text{\cite{Belle:2006pxp}},
%\end{align*}
Using a dataset of $362\;\text{fb}^{-1}$ collected between 2019 and 2021, Belle~II obtained the following results:
\begin{align*}
    S^{\rm Belle\ II}_{\eta'K^0_S}      &= \phantom{-} 0.67 \pm 0.10 \pm 0.04~\text{\cite{Belle-II:2024xzm}}, \\
    S^{\rm Belle\ II}_{K^0_S\pi^0}      &= \phantom{-} 0.75^{+0.20}_{-0.23}\pm 0.04~\text{\cite{Belle-II:2023grc}}, \\
    S^{\rm Belle\ II}_{\phi K^0_S}      &= \phantom{-} 0.54 \pm 0.26^{+0.06}_{-0.08}~\text{\cite{Belle-II:2023uql}}, \\
    S^{\rm Belle\ II}_{K^0_SK^0_SK^0_S} &=          -  1.37^{+0.35}_{-0.45}\pm 0.03~\text{\cite{Belle-II:2024tqu}}, \\
    %,
\end{align*}
where the first uncertainty is statistical and the second is systematic. 
Using the same dataset, Belle~II also performed a measurement of the $CP$-violation parameters in $B^0\to K^{*0}\gamma$ decays. In those decays, also mediated by a one-loop transition in the SM, mixing-induced $CP$ violation is suppressed by the polarisation of the photon and $S$ is expected to be close to $0$. Belle~II finds
\begin{align*}
    S^{\rm Belle\ II}_{K^{*0}\gamma}    &= \phantom{-} 0.00^{+0.27}_{-0.26}\pm 0.03.
\end{align*}
All the above results are compatible with the SM expectation and with previous Belle and BaBar measurements~\cite{HFLAV:2022esi}. 
The results for the $B^0\to K^0_S\pi^0$ decay and the $B^0\to K^{*0}(\to K^0_S\pi^0)\gamma$ decay are more precise than those of Belle\cite{Belle:2006pxp,Belle:2008kbm}, in spite of the smaller dataset used.
This is mainly achieved thanks to an improved $K^0_S$ and $\pi^0$ selection based on machine-learning algorithms.
The decay-time distributions for these two decays are shown in Fig.~\ref{fig:BelleIIdt}.

In addition, Belle~II also introduced a new flavour-tagging algorithm based on a graph-neural-network  which improves the effective tagging efficiency by 18\% with respect to the previous algorithm~\cite{Belle-II:2024lwr}.
The $C\!P$-violating parameters in $B^0 \to J/\psi K^0_S$ decays are measured with this new algorithm to be
\begin{align*}
S^{\rm Belle~II}_{J/\psi K^0_S} &= \phantom{-} 0.724 \pm 0.035 \pm 0.014, \\
C^{\rm Belle~II}_{J/\psi K^0_S} &=          -  0.035 \pm 0.026 \pm 0.013,
\end{align*}
where the first uncertainty is statistical and the second is systematic.
The results are consistent with the world average and the new LHCb result.

\subsection{Time-dependent measurements of the CKM \texorpdfstring{$\gamma$}{} angle}
\label{sec:gamma}

The CKM angle $\gamma$ is defined in Eq.~\eqref{eq:gamma}. The world average is dominated by LHCb measurements of $B^+$, $B^0$ and $B_s^0$ decays, leading to the value of 
$\gamma = (63.8^{+3,5}_{-3.7})^\circ$~\cite{LHCbgamma}. The time-integrated measurements are covered in the summary of WG5~\cite{WG5}. 
A unique way to measure the CKM angle $\gamma$ is via time-dependent measurements of $B_s^0 \to D_s^\mp K^\pm$-like decays. 
Due to the interference between mixing and decay amplitudes, the $CP$-violating parameters in these decays are a combination of the CKM angle $\gamma$ and the relevant mixing phase, namely $\gamma - 2\beta_s$, where $\beta_s$ is defined in Eq.~\eqref{eq:beta_s}. Therefore any measurement from $B_s^0 \to D_s^\mp K^\pm$ decays can be interpreted in terms of either the CKM angle $\gamma$ or $-2\beta_s = \phi_s$ when taking the other parameters from an independent source. 

The LHCb collaboration presented an updated measurement of the CKM angle $\gamma$ using $B_s^0 \to D_s^\mp K^\pm$ decays reconstructed in the Run~2 data-set 
representing $6 \,\mathrm{fb}^{-1}$ of data~\cite{LHCb-gamma}. %collected by the LHCb detector at the centre-of-mass energy of 13 TeV~\cite{LHCb-gamma}. 
The time-dependent asymmetries are defined in Eq.~\eqref{eq:asymmetry}. % replacing the mixing frequency and decay width to the corresponding parameters in the $B_s^0$ system.
As the decay-width difference significantly differs from zero in the 
$B_s$-$\bar B_s$ system, this channel gives access to the hyperbolic terms, allowing for measuring five $CP$-violating observables: $C$, $S$, $\bar{S}$, $\mathcal{A}^{\Delta\Gamma}$ and $\mathcal{\bar{A}}^{\Delta\Gamma}$, where $C = -\bar{C}$. These parameters are related to two final states, either $f=D_s^-K^+$ or $\bar{f} = D_s^+K^-$. 

\begin{figure} [hbt!]
\centering
\includegraphics[width=0.8\textwidth]{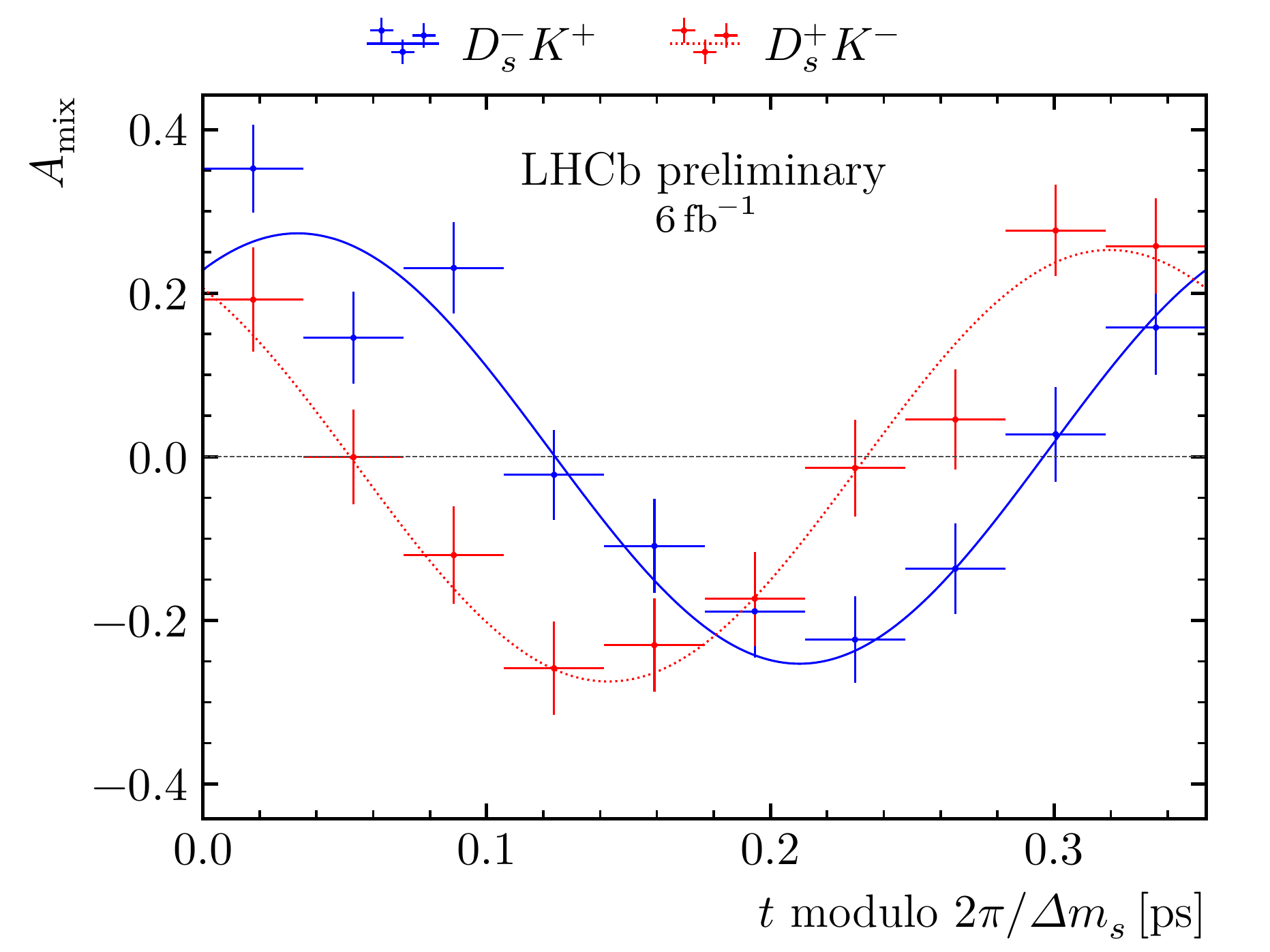}
 \caption{$CP$ asymmetry for the (blue) $D_s^-K^+$ and the (red) $D_s^+K^-$ 
final states, folded into one mixing period,  $2\pi/\Delta m_s$\cite{LHCb-gamma}.}
\label{fig:lhcb-dsk}
\end{figure}

The values of $CP$-violating parameters are measured to be: 
\begin{align*}
C_{D_sK}^{\text{Run\,2}} &= \phantom{-}0.791 \pm 0.061 \pm 0.022, \\
\mathcal{A}^{\Delta\Gamma,\,\text{Run\,2}}_{D_sK} &= -0.051 \pm 0.134 \pm 0.037\\
\mathcal{\bar{A}}^{\Delta\Gamma,\,\text{Run\,2}}_{D_sK} &= -0.303 \pm 0.125 \pm 0.036,\\
S_{D_sK}^{\text{Run\,2}} &= -0.571 \pm 0.084 \pm 0.023,\\
\bar{S}_{D_sK}^{\text{Run\,2}} &= -0.503 \pm 0.084 \pm 0.025,\\
\end{align*}
where the first uncertainty is statistical and the second is systematic. The $CP$ violation in the interference of mixing and decay, i.e. $S \neq \bar{S}$ is obtained at the level of 8.8$\sigma$.
The interpretation in terms of the physical observables with neglecting penguin-loop diagrams or processes beyond the SM gives: 
\begin{align*}
\gamma^{\text{Run\,2}} &= (74 \pm 11)^\circ, \\
\delta_{D_sK}^{\text{Run\,2}} &= (346.9 \pm 6.6)^\circ, \\
r_{D_sK}^{\text{Run\,2}} &= 0.327 \pm 0.038,
\end{align*}
where uncertainties are combined and $\delta$ denotes the strong phase in this process, while $r_{D_sK}$ is an amplitude ratio defined as $r_{D_sK}=|A(\bar{B}_s^0 \to D_s^-K^+)|/|A(B_s^0 \to D_s^-K^+)|$. The time-dependent asymmetry is shown in Fig.~\ref{fig:lhcb-dsk}. 

The presented measurement is the most precise time-dependent measurement of the CKM angle $\gamma$ to-date and agrees with the world average~\cite{HFLAV:2022esi} and global fits by the CKMFitter~\cite{CKMFitter} and UTFit~\cite{UTFit} collaborations.
The overall agreement of this new measurement with the previous measurement of $B_s^0 \to D_s^\mp K^\pm$ using an independent data set of $3 \,\mathrm{fb}^{-1}$ recorded by the LHCb experiment at centre-of-mass energies of 6 and 7 TeV~\cite{LHCb-gamma-run1} is at the level of $1.3 \,\sigma$.
This small deviation is driven by the high value of $\gamma^{\text{Run\,1}} = (128^{+17}_{-22})^\circ$ obtained in the previous measurement.

In the past, two alternative decay modes have been used to extract the CKM angle $\gamma$ using a time-dependent analysis.
The analysis of $B_s^0 \to D_s^\mp K^\pm \pi^+ \pi^-$ decays based on the $9\,\mathrm{fb}^{-1}$ LHCb data set\cite{LHCb-dskpipi} requires precise understanding of the phase space and possible intermediate resonances.
The angular analysis yields an angle of $\gamma_{D_sK\pi\pi} = (44 \pm 12)^\circ$, while a phase-space integrated approach gives $\gamma_{D_sK\pi\pi} = (44^{+20}_{-13})^\circ$.
Time-dependent measurements using $B^0 \to D^\mp \pi^\pm$ decays suffer from low sensitivity due to the small amplitude ratio $r_{D\pi} \approx 0.02$.
The corresponding measurement using $3 \,\mathrm{fb}^{-1}$ of data recorded by the LHCb experiment at centre-of-mass energies of 6 and 7 TeV~\cite{LHCb-dpi-run1} is used to set limits of $\gamma_{D\pi} \in [5^\circ, \, 86^\circ] \cup [185^\circ, \, 266^\circ]$ at 68\% CL by using external input for the angle $\beta$.

Currently, the LHCb collaboration takes action to combine the two time-dependent measurements of $\gamma$ based on $B_s^0 \to D_s^\mp K^\pm$ decays.
The result will replace the previous stand-alone result using only the $3 \,\mathrm{fb}^{-1}$ data set in upcoming LHCb-wide combinations of the CKM angle $\gamma$.

\subsection{New physics in \texorpdfstring{$B_{q} - \bar{B_{q}}$}{}  in connection with the CKM \texorpdfstring{$\gamma$}{} angle.}

%Mixing processes are ideal places to search for NP effects.
In Ref.~\cite{DeBruyn:2022zhw}, it is explored how much space is left for NP through the current data, performing a careful analysis of the Unitarity Triangle. The choice for the input measurements, in particular on the determination of the apex of the UT, plays a central role in the NP searches. Tensions arise between inclusive and exclusive determinations of the CKM matrix elements $|V_{ub}|$ and $|V_{cb}|$. Separate analyses for these two cases is important to be performed. A third possibility, a hybrid scenario combining the exclusive $|V_{ub}|$ with the inclusive $|V_{cb}|$ value is also studied \cite{DeBruyn:2022zhw}. Sizeable differences are observed between the three cases, indicating that it is essential for the puzzles between the different determinations of the CKM parameters to finally be resolved.

In the method that Ref.~\cite{DeBruyn:2022zhw} chooses to follow for the extraction of the UT apex in the SM determination, only  two observables are used: the UT side $R_b$ and the UT angle $\gamma$.
The two parameters are experimentally measured using tree-level decays, and hence  possible NP contamination is expected to be minimal. 
The side $R_b$ is defined as
\begin{equation}\label{eq:Rb}
    R_b \equiv \left(1-\frac{\lambda^2}{2}\right)\frac{1}{\lambda}\left|\frac{V_{ub}}{V_{cb}}\right|
    = \sqrt{\bar\rho\,^2 + \bar\eta\,^2}\:,
\end{equation}
and hence it strongly depends on the determination of the CKM matrix elements $|V_{cb}|$ and $|V_{ub}|$. 

In addition of  utilising purely tree $B\to DK$ decays, the CKM angle $\gamma$  can also be determined using $B\to\pi\pi$, $\rho\pi$, $\rho\rho$ modes and isospin relations. These modes are usually interpreted in terms of the UT angle $\alpha$,
from which $\gamma$ can be extracted using the relation $\gamma=\pi-\alpha-\beta$. Despite the very different dynamics between the corresponding decays, excellent agreement has been found between the two determinations. Although such an agreement is expected in the SM, discrepancies may arise due to NP contributions to the decay amplitudes.

To quantify the impact of different NP scenarios in $B_{q}^0 - \bar{B_{q}}^0$ mixing,  the generalised expressions of the mixing parameters is utilised \cite{Ball:2006xx}:
\begin{align}
 \phi_{q} & = \phi_{q}^{\text{SM}} + \phi_{q}^{\text{NP}} = \phi_{q}^{\text{SM}}  + \arg\left(1 + \kappa_{q} e^{i\sigma_{q}}\right)\:, \label{eq:NP_mix_phiq}\\
    %%%
    \Delta m_{q} & = \Delta m_{q}^{\text{SM}} \left|1 + \kappa_{q} e^{i\sigma_{q}}\right|\:. \label{eq:NP_mix_Dmq} \end{align}
Here, $\kappa_q$ describes the size of the NP effects and $\sigma_q$ is a complex phase accounting for additional $CP$-violating effects. This parametrisation is model independent. 

We study two NP scenarios. Scenario I is the most general one and uses as input $R_b$ and $\gamma$, assuming that there is no NP contribution in these input parameters.  The extraction of the parameters $\kappa_{q}$ and $\sigma_{q}$,  is performed separately for the $B_d^0$ and the $B_s^0$ system. In Scenario II, we use a Flavour Universal New Physics (FUNP) assumption. Here, the assumption is that  $(\kappa_d, \sigma_d) = (\kappa_s, \sigma_s)$, which means that the NP contributions are equal in the $B_d^0$ and the $B_s^0$ system. The determination of the UT apex does not rely on $\gamma$ now but in the sides $R_b$ and $R_t$ instead. 

\begin{figure}[t!]
    \centering
    \includegraphics[width=0.75\textwidth]{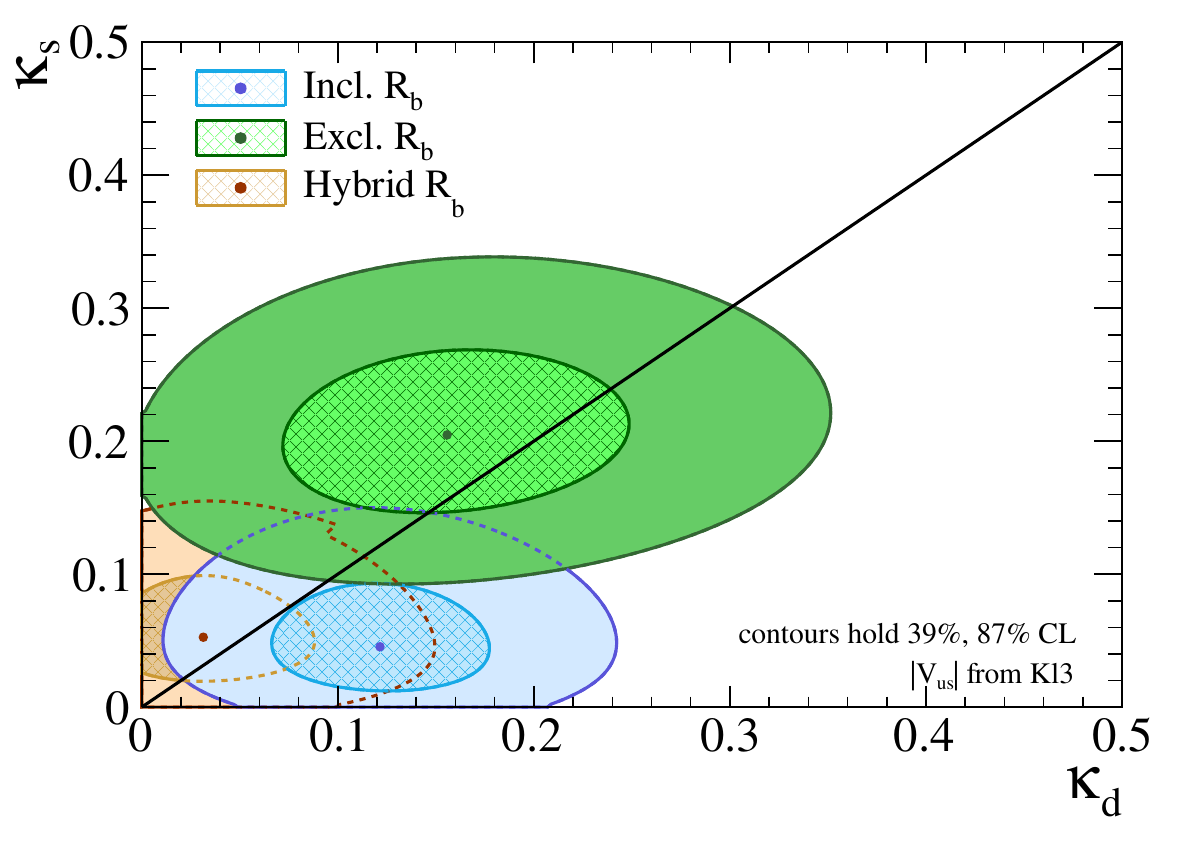}
    \caption{ Fit for the NP parameters $\kappa_d$ and $\kappa_s$, comparing the inclusive, exclusive and hybrid scenarios. The black diagonal line denotes the case $\kappa_d = \kappa_s$, which is the FUNP assumption. The plot is taken from Ref.~\cite{DeBruyn:2022zhw}}
    \label{fig:NP_comparisons_kappa}
\end{figure}

Comparing Scenarios I and II provides a test of whether the data are compatible with the FUNP assumption, and characterizes how the FUMP assumption constrains the NP parameter space. 
For each scenario, three different determinations of $R_b$ are compared using exclusive, inclusive or hybrid inputs for $V_{cb}$ and $V_{ub}$.
Fig.\ \ref{fig:NP_comparisons_kappa} illustrates the correlation between $\kappa_d$ and $\kappa_s$. All three $R_b$ solutions are compatible with the FUNP assumption $\kappa_d = \kappa_s$. 

These NP results have also interesting applications in the analysis of leptonic rare decays $B_q^0\to \mu^+\mu^-$. Key observable is the corresponding SM branching ratio. The branching ratio requires information on $|V_{ts}|$, which is determined through the $|V_{cb}|$ matrix element. It is shown that the dependence on $|V_{cb}|$ can be canceled in the SM by creating the following ratio with the $B_s$ mass difference $\Delta m_s$  \cite{Buras:2003td,Buras:2021nns,Bobeth:2021cxm}:
\begin{equation}
    \mathcal{R}_{s\mu} \equiv {\bar{\mathcal{B}}(B_s\to\mu^+\mu^-)}/{\Delta m_s} \:,
\end{equation}
where possible NP contributions to $B^0_q$-$\bar{B}^0_q$ mixing can be taken into account, following the strategy proposed in \cite{DeBruyn:2022zhw}.

This NP analysis is extrapolated to the high-precision era and future projections are made \cite{DeBruyn:2022zhw}. In the $B_d^0$-$\bar B_d^0$ system, the UT apex is a limiting factor. This is however not the case for the $B_s^0$-$\bar B_s^0$ system, making it a good candidate for searching for new sources of NP.

New opportunities are expected in  NP searches, both related to the determination of the angle $\gamma$ and of the $B^0_q$-$\bar{B}^0_q$ mixing parameters. Future improvements of the precision on $\gamma$ are especially important, given that an improved precision might reveal significant deviations between the different determinations of $\gamma$ due to NP contributions.

\subsection{Determination of \texorpdfstring{$\gamma$}{} from \texorpdfstring{$B_{(s)}\to hh$}{} decays}
Charmless two-body $B_{(s)}\to hh$ decays, where $h= \pi,K$, are important probes of $CP$ violation. Among these decays, the penguin-dominated $B_s^0\to K^-K^+$ decay can also be used to determine the angle $\gamma$ and the $B_s^0$--$\bar{B}_s^0$ mixing phase $\phi_s$ \cite{Fleischer:1999pa,Fleischer:2007hj, Fleischer:2010ib,Ciuchini:2012gd, Fleischer:2016ofb,Fleischer:2016jbf,Nir:2022bbh,Fleischer:2022rkm}. The LHCb collaboration recently observed $CP$ violation in this decay for the first time \cite{LHCb:2020byh}, allowing for the determination of either $\gamma$ or $\phi_s$ from a loop-topology dominated decay \cite{Fleischer:2022rkm}. These penguin modes are very sensitive to effects of new particles. Therefore it is interesting to compare $\gamma$ or $\phi_s$ extracted from these decays with determinations from tree-level dominated decays. 

Looking at the differences between the direct $CP$ asymmetries of several newly measured $B_{(s)}\to hh$ modes by the LHCb collaboration also reveals an interesting pattern. Comparing the modes that only differ by a spectator quark, leads to \cite{Fleischer:2022rkm}
\begin{align}
	\mathcal{A}_{\rm CP}^{\rm dir}(B_s^0 \to K^- K^+) - \mathcal{A}_{\rm CP}^{\rm dir}(B_d^0 \to \pi^- K^+) &= 0.089 \pm 0.031 \ , \nonumber \\
	\mathcal{A}_{\rm CP}^{\rm dir}(B_d^0 \to \pi^- \pi^+) - \mathcal{A}_{\rm CP}^{\rm dir}(B_s^0 \to K^- \pi^+) &= -0.095 \pm 0.040. \label{eq:aCPdirDiff2}
\end{align}
which is quite striking as the difference from zero is at the $(2-3)\sigma$ level. This difference is quite difficult to explain through new-physics effects, because the above decays only differ through their spectator quarks. On the other hand, the $B_s^0 \to K^- K^+$ and $B_d^0 \to \pi^- \pi^+$ modes also have contributions from Exchange ($E$) and penguin-annihilation ($PA$) topologies while the $B_d^0 \to \pi^- K^+$ and $B_s^0 \to K^- \pi^+$ decays do not. This raises the question if the differences in Eq.~\eqref{eq:aCPdirDiff2} can be accommodated by reasonable $E$ and $PA$ contributions. In Ref.~\cite{Fleischer:2022rkm}, both the determination of $\gamma$ and $\phi_s$ from these decays and a new strategy to determine the size of the $E$ and $PA$ contributions was presented. In this Section, these findings are reviewed. 

\vspace{0.2cm}
\noindent {\bf Extracting $\gamma$ using flavour symmetries} 
The $B_s^0 \to K^- K^+$ and $B_d^0 \to \pi^- \pi^+$ decays form a $U$-spin system, with tree ($T$), QCD penguin ($P$) and $E$ and $PA$ contributions. Their amplitudes are parametrised in terms of the penguin parameters $d^{(')}$ and their $CP$-conserving phase $\theta^{(')}$, where the prime indicates the $b\to s$ transition through \cite{Fleischer:2016ofb}  
\begin{align}
    A(B_s^0\to K^+ K^-) &= \sqrt{\epsilon}e^{i\gamma}\mathcal{C}^\prime \left[1 + \frac{1}{\epsilon}d' e^{i\theta'} e^{-i\gamma} \right]\ , \\
    A(B_d^0\to \pi^+\pi^-) &= e^{i\gamma}\mathcal{C}(1 + d e^{i\theta} e^{-i\gamma}) \ ,
\end{align}
where $\epsilon \equiv \lambda^2/(1-\lambda^2)$, with $\lambda$ being the Wolfenstein parameter of the CKM matrix. % normalization is given by
%\begin{equation}\label{eq:Cintro}
 %   \mathcal{C} \equiv \lambda^3 A R_b \left[T+E + P^{(ut)} + PA^{(ut)}\right] \ 
%\end{equation}
The penguin parameter is
\begin{equation} \label{eq:dthetaDef}
    d e^{i\theta}\equiv \frac{1}{R_b}\left[\frac{P^{(ct)} + PA^{(ct)}}{T+E+P^{(ut)}+PA^{(ut)}}\right]\ ,
\end{equation}
with a similar expression for $d'$. In addition, $P^{(qt)} \equiv P^{(q)} - P^{(t)}$, and in analogy for the $PA^{(qt)}$ contribution and $\mathcal{C}$ parametrises the $T+E+P^{(ut)}+PA^{(ut)}$ amplitudes. 
In the $U$-spin limit, the penguin parameters are related via $d e^{i\theta}=d'e^{i\theta'}$ \cite{Fleischer:1999pa}.

\begin{figure}[t]
	\centering 
 	\includegraphics[width=0.6\textwidth]{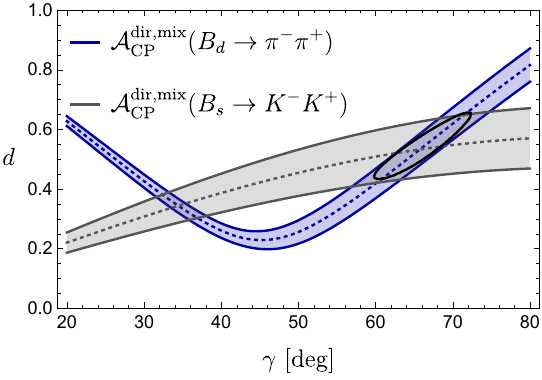}
	\caption{Dependence of the $CP$ asymmetries of the $B_d \to \pi^- \pi^+$ and $B_s \to K^- K^+$ modes on the penguin parameter $d$ and the CKM angle $\gamma$ using data from \cite{LHCb:2020byh}. The overlap region gives the determination of $\gamma$. Plot taken from \cite{Fleischer:2022rkm}.}
	\label{fig:det}
\end{figure}

The direct and mixing-induced $CP$ asymmetries of the $U$-spin system, depend only on $d^{(')},\theta^{(')}$ and $\gamma$, $\phi_s$ and $\phi_d$. Assuming $U$-spin symmetry and taking $\phi_s$ and $\phi_d$, the penguin parameter $d$ and $\gamma$ simultaneously using only the $CP$ asymmetries and by taking $\phi_s$ and $\phi_d$ as inputs. The current data in \cite{LHCb:2020byh} give the contours in the $d$-$\gamma$ plane given in Fig.~\ref{fig:det}. The intersection of the two curves gives
\begin{equation}
    d = d' = 0.52_{-0.09}^{+0.13} \ 
    \end{equation}
and
\begin{equation}\label{eq:ourgam}
    \gamma = (65_{-5}^{+7})^\circ  \ .
\end{equation}
Comparing with the CKM angle $\gamma$ determinations from tree-level decays described in Sec.~\ref{sec:gamma}, especially not considering the determinations from $B_s^0$ decays leads to impressive agreement.

\vspace{0.2cm}
\noindent {\bf Probing the Exchange and Penguin annihilation contributions} 
The new LHCb data revealed a large difference between the direct $CP$ asymmetries of $B_d^0\to \pi^- \pi^+, B_s^0 \to K^- K^+$ system and the $B_d^0 \to \pi^- K^+, B_s^0 \to K^- \pi^+$ system in Eq.~\eqref{eq:aCPdirDiff2}. Within the SM, this difference can only be explained through (sizeable) exchange ($E$) and penguin-annihilation ($PA$) contributions, which contribute to the first $U$-spin system but not to the second. These contributions are denoted by the hadronic parameters $x^{(\prime)}$ and $r_{PA}^{(\prime)}$, respectively. The mismatch between the two $U$-spin systems is given by 
\begin{equation} \label{eq:zetapdef}
	\zeta' \equiv |\zeta'| e^{i \omega'} \equiv \frac{1+x'}{1+r_{PA}'} 
\end{equation}
with
\begin{equation} \label{eq:xrpadef}
	x^{(\prime)} \equiv |x^{(\prime)}| e^{i \sigma^{(\prime)}} \equiv \frac{E^{(\prime)} + PA^{(ut)(\prime)}}{T^{(\prime)} + P^{(ut)(\prime)}}, \quad r_{PA}^{(\prime)} \equiv |r_{PA}^{(\prime)}| e^{i \theta_{PA}^{(\prime)}} \equiv \frac{PA^{(ct)(\prime)}}{P^{(ct)(\prime)}}.
\end{equation}
Figure~2 in Ref.~\cite{Fleischer:2022rkm}, describes in detail a new strategy to determine these parameters from the experimental data. An brief outline of this new strategy is presented below.

\noindent $\bullet$ {\bf{Step 1}}: Using instead $\gamma$ as an input, allows to determine the hadronic parameters defined in Eq.~\eqref{eq:dthetaDef} by assuming the $U$-spin symmetry in the $B_d^0 \to \pi^- \pi^+, B_s^0 \to K^- K^+$ system. This gives
\begin{equation}\label{eq:step1}
    d = d' =  0.39 \pm 0.05 , \qquad \theta = \theta' =  (140 \pm 5)^\circ  \ ,
\end{equation}
where compared to the above $d$ is shifted down by $1\, \sigma$ level. 

\vspace{0.2cm}
\noindent $\bullet$ {\bf{Step 2}}:   
Similarly, the analogous penguin parameter $\tilde{d}$ can be obtained from the $CP$ asymmetries of the $B_s \to K^-\pi^+$ and $B_ds \to \pi^-K^+$decays. Comparing gives
\begin{equation}\label{eq:step2}
    |\zeta| \equiv \tilde{d}/d  = 1.3\pm 0.2\ , \quad \omega\equiv \tilde{\theta}-\theta =(16 \pm 5)^\circ \ ,
\end{equation}
which measure contribution of the $E^{(\prime)}$ and $PA^{(\prime)}$ topologies in the $B_d^0 \to \pi^- \pi^+$ ($B_s^0 \to K^- K^+$) decay with respect to the tree and penguin amplitudes. From the values in Eq.~\eqref{eq:step2}, it is concluded in \cite{Fleischer:2022rkm} that the difference in the direct $CP$ asymmetries in Eq.~\eqref{eq:aCPdirDiff2} can be accommodated by exchange
and penguin annihilation effects at the level of (20–30)\%. This size lies in the range of what is expected theoretically for such contributions and does not signal any anomalously enhanced rescattering or long-distance effects. 

\begin{figure}
	\centering
	\subfigure{\includegraphics[height=0.3\textwidth]{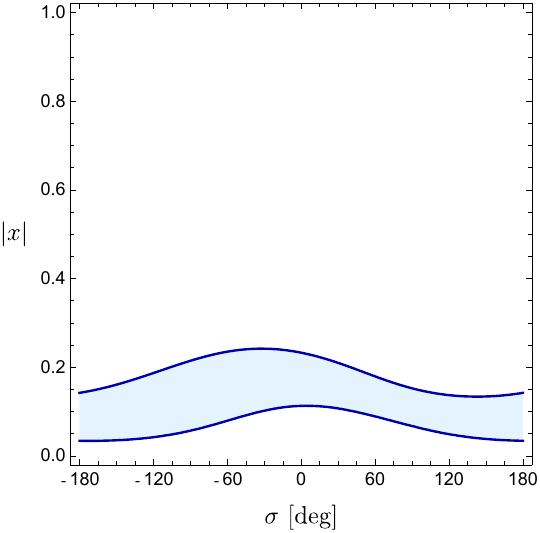}}
	\subfigure{\includegraphics[height=0.3\textwidth]{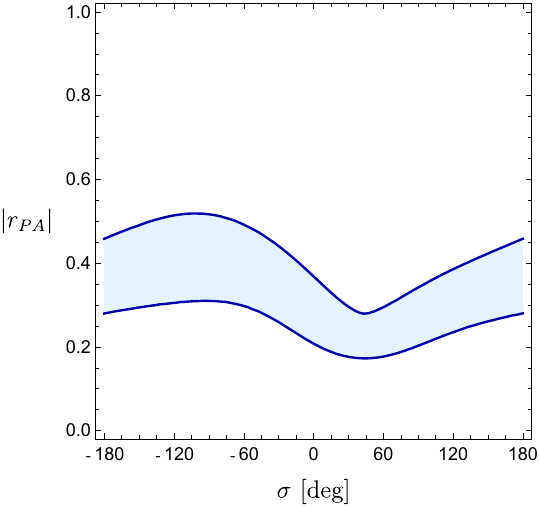}}
	\subfigure{\includegraphics[height=0.3\textwidth]{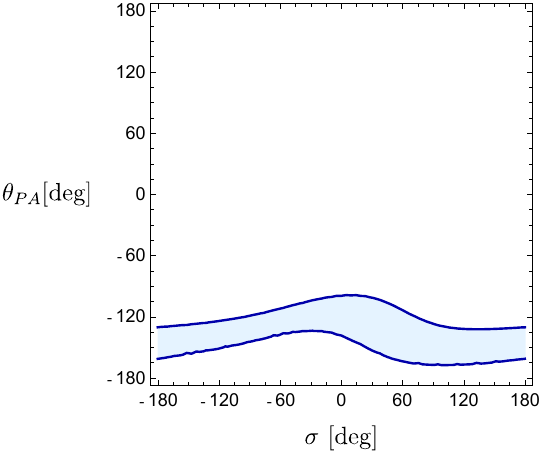}}
	\caption{Extracted $1\, \sigma$ allowed space for the exchange ($x, \sigma$) and penguin-annihilation ($r_{PA}, \theta_{PA}$) parameters. Figure taken from \cite{Fleischer:2022rkm}.}
	\label{fig:EPAfitRegions}
\end{figure}

\vspace{0.2cm}
\noindent $\bullet$ {\bf{Step 3}}: To determine $x$ and $r_{PA}$ from $\zeta$ requires additional information. This can be provided by the $B_s^0 \to \pi^- \pi^+, B_d^0 \to K^- K^+$ $U$-spin system, which receives only contributions from $E$ and $PA$ topologies. At the moment, only the branching ratios of these decays are measured. In Ref.~\cite{Fleischer:2022rkm} the constraints Fig.~\ref{fig:EPAfitRegions} were obtained. In the future, possible measurements of the $CP$ asymmetries in the $B_s^0 \to \pi^- \pi^+, B_d^0 \to K^- K^+$ system would allow pinning down these contributions even further.

\vspace{0.2cm}
\noindent {\bf Extracting $\phi_s$ from penguin decays} 
Finally, the penguin-dominated $B_s^0 \to K^-K^+$ decay mode can also be used to determine the $B_s^0$--$\bar{B}_s^0$ mixing phase $\phi_s$. Doing so requires taking $\gamma$ as an input. 

The $CP$ asymmetries of $B_s^0\to K^- K^+$ allow the extraction of the effective phase
\begin{equation}
    \sin\phi_s^{\rm eff} = \frac{ \mathcal{A}_{\rm CP}^{\rm mix}(B_s^0\to K^-K^+)}{ \sqrt{1-\left[\mathcal{A}_{\rm CP}^{\rm dir}(B_s^0\to K^-K^+)\right]^2}} \ .
\end{equation}
Using the LHCb data for the $CP$ asymmetries from \cite{LHCb:2020byh}, gives $\phi_s^{\rm eff} = -(8.1 \pm 1.9)^\circ$ which is defined as
\begin{equation}
    \phi_s^{\rm eff} \equiv \phi_s + \Delta \phi_{KK} \ ,
\end{equation}
where the parameter $\Delta\phi_{KK}$ is a hadronic phase shift
\begin{equation}\label{eq:deltaphikk}
    \tan\Delta \phi_{KK} = 2 \epsilon \sin\gamma \left[\frac{d' \cos\theta'+ \epsilon \cos\gamma}{d^{\prime 2}+ 2 \epsilon d' \cos\theta' \cos\gamma + \epsilon^2 \cos 2\gamma}\right] \ ,
\end{equation}
which depends on the hadronic parameters $d',\theta'$ discussed before. In~\cite{Fleischer:2016ofb} and \cite{Fleischer:2022rkm}, two strategies to determine $\Delta\phi_{KK}$ using semileptonic decays were presented:

\underline{Strategy I:}
This strategy uses (double) ratios of non-leptonic and semileptonic $B_{(s)}$ decay rates \cite{Fleischer:2016ofb}: 
\begin{equation}\label{eq:defRpi}
	R_{\pi}  \equiv  \frac{ \Gamma(B_d^0\to \pi^-\pi^+)}{|d\Gamma(B^0_d\rightarrow \pi^- \ell^+ 
	\nu_\ell)/dq^2|_{q^2=m_\pi^2}} \ , 	\quad\quad R_{K}  \equiv  \frac{ \Gamma(B_s^0\to K^- K^+)_{\rm theo}}{|d\Gamma(B^0_s\rightarrow K^- \ell^+ 
	\nu_\ell)/dq^2|_{q^2=m_K^2}} \ .
\end{equation}
In terms of the hadronic parameters, $R_\pi$ is proportional to 
\begin{equation}\label{eq:defrpi}
r_\pi \equiv 1+d^2-2d\cos\theta\cos\gamma		 \ ,
\end{equation}
while $R_K$ is proportional to 
\begin{equation}\label{eq:defrk}
    r_K \equiv 1+\left(\frac{d^{\prime}}{\epsilon}\right)^2+2 \frac{d^\prime}{\epsilon} \cos\theta^\prime \cos\gamma .
\end{equation}
Taking then the ratio of $R_\pi$ and ${R}_K$ yields
\begin{equation} \label{eq:rkTh}
    {r}_K = \frac{{R}_K}{R_\pi} \left( \frac{|V_{ud}| f_\pi}{|V_{us}| f_K} \right)^2 \frac{X_\pi}{{X}_K} \left( {\xi}_\text{NF}^a \right)^2 r_\pi \ ,
\end{equation}
where also the decay constants $f_{\pi,K}$ and the ratio of $B_s\to K$ and $B\to \pi$ form factors enter through $X_K$ and $X_\pi$, respectively (see (56) in Ref.~\cite{Fleischer:2022rkm}). These inputs are precisely determined, and the remaining theoretical uncertainty comes from the ratio
\begin{equation}\label{eq:xidef}
    {\xi}_\text{NF}^a \equiv \left| \frac{1 + r_P}{1 + {r}'_P} \right| \left| \frac{1 + x}{1+x'} \right| \left| \frac{a_{\rm NF}^T}{{a}_{\rm NF}^{T'}} \right|,
\end{equation}
which parametrises the non-factorisable $U$-spin-breaking contributions. Here $r_P = P^{(ut)}/T$ and $a_{\rm NF}^T$ describes the non-factorisable contributions to the colour-allowed tree topology, which can be obtained using QCD factorisation.  
% \begin{equation} \label{eq:defaNFd}
% 	a_{\rm NF} \equiv (1+r_P)(1+x) a^T_{\rm NF} \ ,
% \end{equation}
% where $r_P = P^{(ut)}/T$ denotes the ratio of penguin to tree topologies. Finally, $a_{\rm NF}^T$ describes the non-factorisable contributions to the colour-allowed tree topology, which can be computed in QCD factorisation \cite{Beneke:1999br,Beneke:2001ev}.
In the exact $U$-spin symmetry, $\xi_{\rm NF}^a=1$, while potential $U$-spin-breaking corrections only enter through the double-ratio in $\xi$. 

In Ref.~\cite{Fleischer:2022rkm}, the uncertainty on $\xi_{\rm NF}^a$ was re-evaluated, using the results of Fig.~\ref{fig:EPAfitRegions}. This also allowed to check the assumptions made in Ref.~\cite{Fleischer:2016ofb}. This led to: 
% \begin{equation}
%     \Xi_x \equiv \left|\frac{1+x}{1+x'}\right| = 1 + x \xi_x + \mathcal{O}(x^2) \ ,
% \end{equation}
% where $\Xi_x$ measures the $U$-spin-breaking corrections in the exchange topologies. Including $U$-spin-breaking corrections at the level of $20\%$ yields with a conservative estimate of $x \sim 0.1\pm 0.1$ from Fig.~\ref{fig:EPAfitRegions}, we find a $4\%$ uncertainty from $\Xi_x$. To estimate the remaining uncertainty on $\xi_{\rm NF}^a$, we follow our previous analysis \cite{Fleischer:2016ofb} in which we found a $1\%$ uncertainty from $a_{\rm NF}^T$ and a $5\%$ uncertainty from the $r_P$ part. The latter requires also information from the charged $B^+$ modes and we explicitly checked that the estimates we made in \cite{Fleischer:2016ofb} are in agreement with current data. Adding these uncertainty contributions in quadrature, we find 
\begin{equation}\label{eq:xiin}
    \xi_{\rm NF}^a = 1.00 \pm 0.07 \ ,
\end{equation}
which results in a theoretical uncertainty of only $0.8^\circ$ on $\Delta\phi_{KK}$. Recently this decay has been observed for the first time by the LHCb collaboration \cite{LHCb:2020ist}. However, for this strategy the differential semileptonic $B_s^0 \to K^- \ell^+ \nu_\ell$ decay rate at $q^2=m_K^2$ is required as input, which is currently not available. Therefore, this strategy cannot be fully implemented at this moment.

\underline{Strategy II:}
An alternative strategy is to use only the nonleptonic $B_s^0\to K^-K^+$ and $B_d^0\to \pi^-\pi^+$ rates and the ratio:
\begin{align} \label{eq:kObservable}
    K &\equiv \frac{1}{\epsilon} \left[\frac{m_{B_s}}{m_{B_d}} \frac{\Phi(m_\pi/m_{B_d},m_\pi/m_{B_d})}{\Phi(m_K/m_{B_s},m_K/m_{B_s})} \frac{\tau_{B_d}}{\tau_{B_s}}\right] \frac{\mathcal{B}(B_s^0 \to K^- K^+)_{\rm theo}}{\mathcal{B}(B_d^0 \to \pi^-\pi^+)} %= 105.3 \pm 9.6 
    \nonumber \\
&=  \left|\frac{\mathcal{C}^\prime}{\mathcal{C}}\right|^2\frac{1 + 2(d'/\epsilon) \cos\theta' \cos\gamma + (d'/\epsilon)^2}{1 - 2 d \cos\theta \cos\gamma + d^2} ,
\end{align}
 where $\Phi$ is a phase space function. The experimental data from LHCb \cite{LHCb:2020byh} gives $K=105.3\pm 9.6$. To use this value to extract the penguin parameters requires a theoretical determination of $\mathcal{C}/\mathcal{C}'$, which can be estimated through 
\begin{equation}\label{eq:cfact}
    \left|\frac{\mathcal{C}}{\mathcal{C}'}\right| = \frac{f_\pi}{f_K} \left[\frac{m_{B_d}^2 - m_\pi^2}{m_{B_s}^2 - m_K^2}\right] \left[\frac{F_0^{B_d \pi}(m_\pi^2)}{F_0^{B_s K}(m_K^2)}\right] \xi^a_{\rm NF}\ ,
\end{equation}
where the form factors are known from \cite{Khodjamirian:2017fxg}. Finally, using also $\gamma$ as input and $\xi_{\rm NF}^a$ from Eq.~\eqref{eq:xiin}, gives
\begin{equation}
    \Delta\phi_{KK} = -(4.5 \pm 5.3)^\circ \ , \quad\quad  \phi_s = -(3.6 \pm 5.7)^\circ \ .
\end{equation}
The uncertainty on $\phi_s$ is dominated by the experimental uncertainties on $B_s^0\to K^-K^+$ $CP$ asymmetries and those on the form factors. 

This new determination is in good agreement with the $\phi_s$ obtained from $B_s^0\to J/\psi \phi$ and agrees with the SM predictions. However, more interestingly, it also leaves significant room for new physics.

In the future, with more precise $CP$ asymmetries, it will be interesting to see if the value of $\phi_s$ from the penguin-modes and that from $B_s^0\to J/\psi \phi$ analyses remain in agreement. If a discrepancy would arise, this signifies new $CP$-violating physics, to which especially the penguin-dominated determination discussed above is sensitive.

\section{Determination of the \texorpdfstring{$\Delta \Gamma$}{} width difference}

\begin{figure} [hbt!]
\centering
\includegraphics[width=0.75\textwidth]{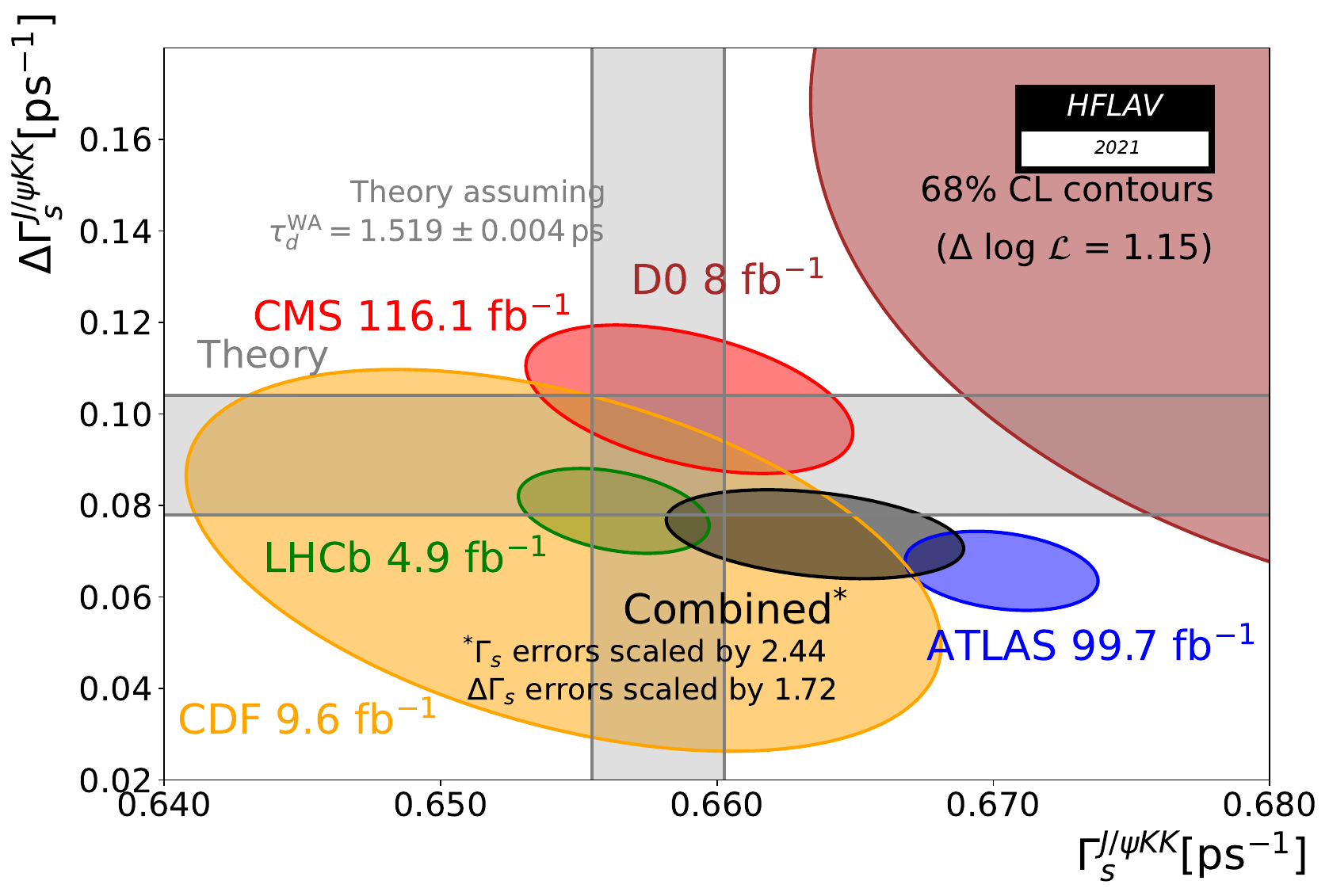}
 \caption{Experimental results of $\Gamma_s$ vs $\Delta \Gamma_s$ from $B_s^0 \to J/\psi KK$ decays averaged by HFLAV group~\cite{HFLAV:2022esi}.}
\label{fig:GsvsDGs}
\end{figure}

In the $B^0_s$ system estimation the decay-width difference $\Delta \Gamma_s$ varies in the range (7.6-9.2)$\times$10$^{-2}$ ps$^{-1}$ depending on the renormalisation choice~\cite{Asatrian,HPQCD,Lenz,Gerlach}. This quantity is measured in channels such as $B_s^0 \to J/\psi\phi$~\cite{ATLASDGs,CMS-phis,LHCb-phis}. 
The experimental results are in tension as shown in Fig.~\ref{fig:GsvsDGs}. Therefore new studies from  experimental and theoretical sides are needed.   

\subsection{New results from LHCb}

\begin{figure} [hbt!]
\centering
\includegraphics[width=0.8\textwidth]{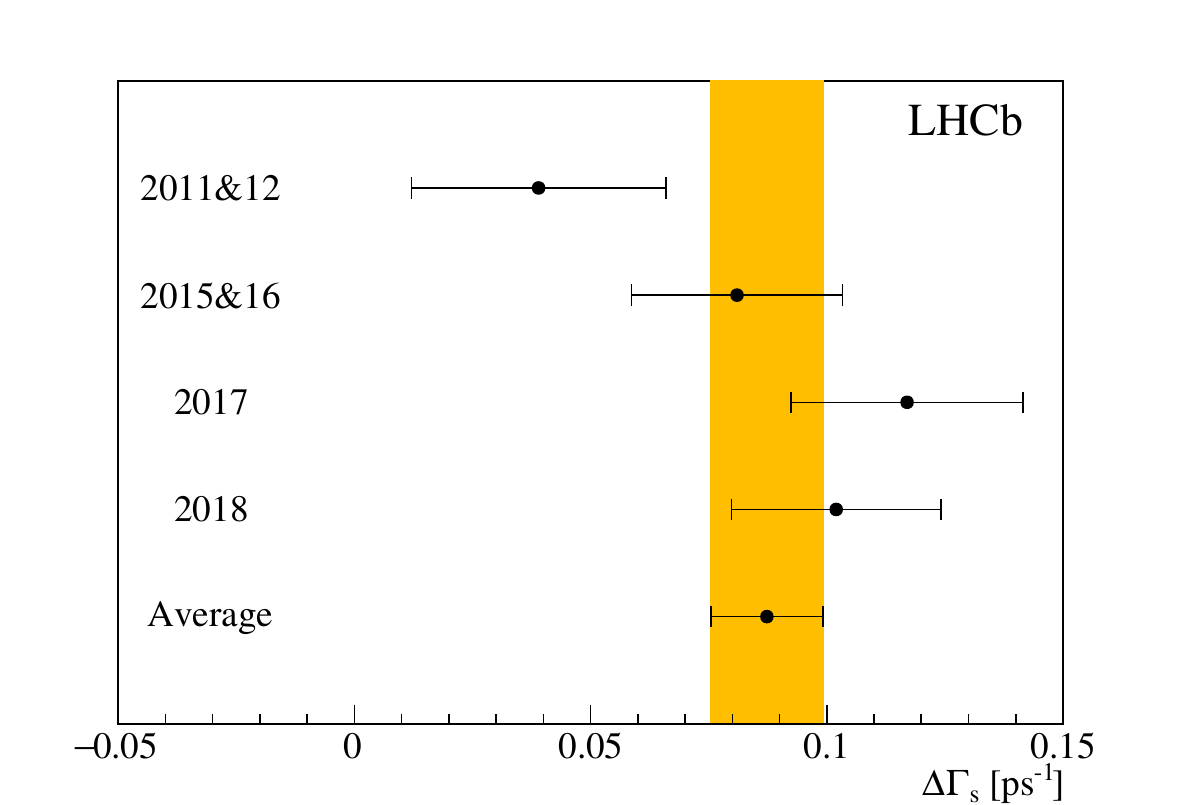}
 \caption{Measurement of $\Delta \Gamma_s$ 
 for the four datasets (2011-2012, 2015-2016, 2017 and 2018) and their weighted average. The orange band corresponds to 1 standard deviation. Taken from \cite{LHCbGDsNew}.}
\label{fig:lhcb-dgs}
\end{figure}

The LHCb experiment presented for the first time measurement on the $\Delta \Gamma_s$ parameter exploiting new decays: $B_s^0 \to J/\psi\eta'$ and $B_s^0 \to J/\psi\pi^+\pi^-$~\cite{LHCbGDsNew}.
The analysis uses the full LHCb Run\,1 and Run\,2 data, corresponding to 9 fb$^{-1}$ of integrated luminosity.
The width difference, $\Delta \Gamma_s$, is obtained from the decay-width difference between a $CP$-odd and a $CP$-even $B_s^0$ mode. It is possible thanks to the small value of $\phi_s$, which implies that with a good approximation $CP$-even modes determine the light mass eigenstate lifetime ($\tau_{L} = 1/\Gamma_{L}$), while $CP$-odd corresponds to the heavy mass eigenstate lifetime $\tau_{H} = 1/\Gamma_{H}$. The measurement is performed in bins of decay time using the ratio of yields in given intervals ($t_1$,$t_2$): 
\begin{equation*}
R_{i} = \frac{N_{L}}{N_{H}} \propto  \frac{[e^{-\Gamma_s t (1+y)}]^{t1}_{t2}}{[e^{-\Gamma_s t (1-y)}]^{t1}_{t2}}\times \frac{1-y}{1+y}
\end{equation*}
where $y=\Delta \Gamma_s/2\Gamma_s$ and $N_{L}$ ($N_{H}$) denotes the yield of  $CP$-even ($CP$-odd) mode in given interval. 
The results are obtained for periods of data-taking as shown in Fig.~\ref{fig:lhcb-dgs}, the combined value of $\Delta \Gamma_s$ yields: 
\begin{equation*}
\Delta \Gamma_s =  0.087 \pm 0.012 \pm 0.009 \quad\rm{ps}^{-1},
\end{equation*}
where the first uncertainty is statistical and second is systematic. 
The value agrees with both HLFAV averages~\cite{HFLAV:2022esi} obtained from the time dependent measurements of $B_s^0 \to J/\psi\phi$ decays, as well as including  constraints from other untagged effective lifetime measurements. 

\subsection{NNLO QCD corrections}
\subsubsection{Calculation overview}
The latest update from on the theoretical calculation of $\Delta\Gamma$ from Refs.~\cite{Gerlach:2022wgb,Gerlach:2022hoj} focuses on reducing the perturbative uncertainties in the leading $\mathcal{O}((\Lambda_\text{QCD}/m_b)^0)$ terms. This is achieved through a matching calculation of a $\lvert\Delta B\rvert = 2$ matrix element calculated within effective $\lvert\Delta B\rvert = 1$ and $\lvert\Delta B\rvert = 2$ theories, where the high-energy and low-energy effects factorise into the matching coefficients and the operator matrix elements respectively. The Wilson coefficients of the first Operator Product Expansion, i.e.~for the $\lvert\Delta B\rvert = 1$ Hamiltonian, have been calculated to three-loop order in previous works \cite{Gambino:2003zm, Gorbahn:2004my, Gorbahn:2005sa}. To obtain only the leading terms in $\Lambda_\text{QCD}/m_b$, the Heavy Quark Expansion (HQE) is used for the transition operator on the $\lvert\Delta B\rvert = 2$ side, which allows us to expand the operators in $\Lambda_\text{QCD}/m_b$ \cite{Khoze:1983yp, Shifman:1985, Khoze:1986fa, Chay:1990da, Bigi:1991ir, Bigi:1992su, Bigi:1993fe, Blok:1993va, Manohar:1993qn, Lenz:2014jha}. The matching calculation is done methodically by first calculating the imaginary part of the $B_s \rightarrow \overline{B}_s$ mixing amplitude in the two effective field theories, renormalising the results and then matching the coefficients of the $\lvert\Delta B\rvert = 2$ operators to the result from the $\lvert\Delta B\rvert = 1$ calculation.

The definition of the Hamiltonian for the $\lvert\Delta B\rvert = 1$ theory in the Chetyrkin-Misiak-Münz basis can be found in~\cite{Chetyrkin:1997gb}. It is worth noting that in the application this basis is particularly useful for automated calculations as it circumvents all of the complications related to $\gamma_5$ in dimensional regularisation.  Another issue related to dimensional regularisation with $d=4-2\epsilon$ dimensions is the appearance of so-called evanescent operators, which are of order $\epsilon$ and vanish in four dimensions due to Fierz identities. However, the evanescent operators mix with physical operators and need to be taken into consideration when renormalising bare amplitudes. 

To calculate the width difference $\Delta\Gamma$, the absorptive part of the scattering matrix element needs to be evaluated, which decomposes into a sum of terms with different CKM factors. This prompts to decompose the $\lvert\Delta B\rvert = 2$ matching coefficients in an analogous fashion. From these considerations, the off-diagonal matrix element of the decay width in the $\lvert\Delta B\rvert = 1$ theory can be written as
\begin{equation}
\Gamma_{12} = \frac{1}{M_B} \sum_{\alpha,\beta} \lambda_\alpha \lambda_\beta \,\text{Im}(\mathcal{M}_{\alpha\beta}),
\end{equation}
and in the $\lvert\Delta B\rvert = 2$ theory as
\begin{equation}
\Gamma_{12} = - \frac{G_F^2 m_b^2}{24 \pi^2 M_B} \sum_{\alpha,\beta} \lambda_\alpha \lambda_\beta \left[H^{\alpha\beta}  \bra{B}Q \ket{\bar{B}} + \tilde{H}_S^{\alpha\beta}  \bra{B} \tilde{Q}_S \ket{\bar{B}} \right] + \mathcal{O}\left(\frac{\Lambda_\text{QCD}}{m_b}\right),\label{eq:dGamma_full}
\end{equation}
where $M_B$ is the mass of the $B$ meson. In the context of the HQE, the matching coefficients $H^{\alpha\beta}$ and $\tilde{H}_S^{\alpha\beta}$ are calculated as expansions in $z \equiv m_c^2/m_b^2$. $Q$ and $\tilde{Q}_S$ are the physical operators of the $\lvert \Delta B \rvert = 2$ transition operator defined in Ref.~\cite{Gerlach:2022wgb}.

As alluded to previously, the low-energy and high-energy physics factorise with the matching coefficients $H^{\alpha\beta}$ and $\tilde{H}_S^{\alpha\beta}$ containing the perturbative high-energy physics that is the main goal of the theoretical calculation described here. The low-energy behaviour captured in the operator matrix elements of the physical operators needs to be extracted from either QCD sum rules \cite{Ovchinnikov:1988zw, Reinders:1988aa, Korner:2003zk, Mannel:2011iqd, Grozin:2016uqy, Kirk:2017juj, King:2019lal, King:2021jsq} or lattice QCD calculations \cite{Davies:2019gnp,Dowdall:2019bea} and is used as an input in the prediction of $\Delta \Gamma$. For more details on the calculation, see Ref.~\cite{Gerlach:2024qlz}.

\subsubsection{Results}
The update on the theoretical calculation in Refs.~\cite{Gerlach:2022wgb, Gerlach:2022hoj} provides new results for the current-current operator contributions at NNLO by combining novel calculations with known results for diagrams with closed charm loops. Contributions from penguin operators are also updated to higher accuracy than in previous calculations. All calculations are done as a naive expansion in $m_c/m_b$ and only the leading term in $z\equiv m_c^2/m_b^2$ is determined, see Tab.~\ref{tab:dGamma_updated_ops} for a summary.
\begin{table}
\centering
\begin{tabular}{@{}  l  l  l @{}}
\toprule
Contribution & Previous results & Refs.\cite{Gerlach:2022wgb, Gerlach:2022hoj} \\ \midrule
$P_{1,2} \times P_{3-6}$ & 2 loops, $z$-exact, $n_f$-part only \cite{Asatrian:2017qaz, Asatrian:2020zxa}  & 2 loops, $\mathcal{O}(z)$, full\\
$P_{1,2} \times P_8$ & 2 loops, $z$-exact, $n_f$-part only \cite{Asatrian:2017qaz, Asatrian:2020zxa}&  2 loops, $\mathcal{O}(z)$, full\\
$P_{3-6} \times P_{3-6}$ & 1 loop, $z$-exact, full \cite{Beneke:1996gn} &  2 loops, $\mathcal{O}(z)$, full\\
$P_{3-6} \times P_{8}$ & 1 loop, $z$-exact, $n_f$-part only \cite{Asatrian:2017qaz, Asatrian:2020zxa}  &  2 loops, $\mathcal{O}(z)$, full\\
$P_{8} \times P_{8}$ & 1 loop, $z$-exact, $n_f$-part only \cite{Asatrian:2017qaz, Asatrian:2020zxa} &  2 loops, $\mathcal{O}(z)$, full\\
$P_{1,2} \times P_{1,2}$ & 3 loops, $\mathcal{O}(\sqrt{z})$, $n_f$-part only \cite{Asatrian:2017qaz, Asatrian:2020zxa} &  3 loops, $\mathcal{O}(z)$, full\\
\bottomrule
\end{tabular}
\caption{Updated contributions to the theoretical value of $\Delta \Gamma$. ``Full'' contributions refers to the fact that both the fermionic and non-fermionic pieces have been calculated.}\label{tab:dGamma_updated_ops}
\end{table}

To improve the numerical accuracy of the result, it is helpful to consider the ratio $\Delta\Gamma/\Delta M$. In this ratio, the dependence on $\lvert V_{cb}\rvert$ drops out and most of the dependence on the bag parameters is also removed. The NNLO corrections from QCD to $\Delta M$ are already known and have been published in Ref.~\cite{Buras:1990fn}. Finally, the theoretical uncertainty also depends on the choice of renormalisation scheme of the mass $m_b^2$ in Eq.~\eqref{eq:dGamma_full} multiplying $\Delta\Gamma$. In a first step, one usually employs a pole mass but subsequently trades it for an MS mass or the potential-subtracted (PS) mass \cite{Beneke:1998rk}, which have better infrared properties. The results for the resulting ratio in the different mass schemes of the overall $m_b^2$ factor are given by
\begin{align}
\left.\frac{\Delta\Gamma_s}{\Delta M_s}\right\rvert_{\overline{\text{MS}}} &= \left ( 4.33 \, \substack{+0.23 \\ -0.44}_\text{scale} \, \substack{+0.09 \\ -0.19}_{\text{scale}, 1/m_b} \pm 0.12_{B\tilde{B}_S} \pm 0.78_{1/m_b} \pm 0.05_\text{input} \right) \times 10^{-3},\\
\left.\frac{\Delta\Gamma_s}{\Delta M_s}\right\rvert_{\text{PS}} &=  \left ( 4.20 \, \substack{+0.36 \\ -0.39}_\text{scale} \, \substack{+0.09 \\ -0.19}_{\text{scale}, 1/m_b} \pm 0.12_{B\tilde{B}_S} \pm 0.78_{1/m_b} \pm 0.05_\text{input} \right) \times 10^{-3},
\end{align}
where the subscripts on the uncertainties indicate their origin \cite{Gerlach:2022wgb}. 
 
Using the experimental value for $\Delta M_s$ \cite{HFLAV:2022esi},
 \begin{equation}
\Delta M_s^\text{exp} =  (17.7656 \pm 0.0057) \;\text{ps}^{-1},
\end{equation}
the theoretical prediction for $\Delta \Gamma_s$ is updated to be
\begin{equation}
\Delta \Gamma_s^\text{th} = (0.076 \pm 0.017) \;\text{ps}^{-1}.
\end{equation}
With this update, the theoretical uncertainty is about three times as large as the experimental uncertainty given in Eq.~\eqref{eq:exp}. Calculations to further improve upon this are already underway and are aiming to improve the accuracy by including higher-order terms in $z$ as well as the penguin operator contributions at NNLO.

\section{Conclusion}
%---------------------------------------------------------------------------------------%
New advancements in the accuracy of mixing and mixing-related $CP$-observables were presented at the 12th Workshop on the CKM Unitarity Triangle. They  provide important insights into the principles governing nature, both in experimental findings and theoretical developments.

$\quad$ \\
{\bf Acknowledgements.} Agnieszka, Thibaud and Vladyslav would like to thank all speakers of the session for their excellent talks. The proceedings authors are also grateful to the organisers for preparing the CKM 2023 workshop so well. 
Agnieszka Dziurda expresses her gratitude to the Ministry of Science and Higher Education in Poland, for financial support under the contract no 2022/WK/03.
Felix Erben has received funding from the European Union’s Horizon Europe research and innovation programme under the Marie Sk\l{}odowska-Curie grant agreement No. 101106913. 
Thomas Latham is supported by the Science and Technology Facilities Council (UK).
Pascal Reeck was supported by the Deutsche Forschungsgemeinschaft (DFG, German Research Foundation) under grant 396021762 - TRR 257 ``Particle Physics Phenomenology after the Higgs Discovery''.

\bibliographystyle{amsplain}

\end{document}